\documentclass[preprint,showpacs,showkeys,preprintnumbers,amsmath,amssymb,prb]{revtex4}
\usepackage{graphicx, epsfig}
\usepackage{amssymb}
\usepackage{subfigure}
\usepackage{sidecap}
\def\a{{\alpha}}
\def\e{{\epsilon}}
\def\bS{{\mathbf{S}}}
\def\beq{\begin{equation}}
\def\eeq{\end{equation}}
\def\s{{\sigma}}
\def\bss{{\bar{\sigma}}}
\def\beqr{\begin{eqnarray}}
\def\eeqr{\end{eqnarray}}
\def\b{{\beta}}
\def\half{{1\over2}}
\def\tlambda{{\tilde{\lambda}}}
\begin{document}

\title{Interactions, superconducting $T_c$, and fluctuation
        magnetization for two coupled dots in the crossover between
        the Gaussian Orthogonal and Unitary ensembles}

\author{Oleksandr Zelyak}
  \email{zelyak@pa.uky.edu}
\author{Ganpathy Murthy}
  \email{murthy@pa.uky.edu}
\affiliation{
Department of Physics and Astronomy, 
University of Kentucky, Lexington, Kentucky 40506, USA
}
\author{Igor Rozhkov}
  \email{rozhkois@notes.udayton.edu}
\affiliation{
Department of Physics, University of Dayton, 300 College Park, Dayton, OH 45469
}

\date{\today}

\begin{abstract}
  We study a system of two quantum dots connected by a hopping
  bridge. Both the dots and connecting region are assumed to be in
  universal crossover regimes between Gaussian Orthogonal and Unitary
  ensembles. Using a diagrammatic approach appropriate for energy
  separations much larger than the level spacing we obtain the
  ensemble-averaged one- and two-particle Green's functions. It turns
  out that the diffuson and cooperon parts of the two-particle Green's
  function can be described by separate scaling functions. We then use
  this information to investigate a model interacting system in which
  one dot has an attractive $s$-wave reduced Bardeen-Cooper-Schrieffer
  interaction, while the other is noninteracting but subject to an
  orbital magnetic field. We find that the critical temperature is
  {\it nonmonotonic} in the flux through the second dot in a certain
  regime of interdot coupling. Likewise, the fluctuation magnetization
  above the critical temperature is also nonmonotonic in this regime,
  can be either diamagnetic or paramagnetic, and can be deduced from
  the cooperon scaling function.
\end{abstract}



\maketitle

\section{Introduction}

The idea of describing a physical system by a random matrix
Hamiltonian to explain its spectral properties goes back to
Wigner\cite{wigner_1955,wigner_1957}.  It was further developed by
Dyson, Mehta and others, and became the basis for Random Matrix Theory
(RMT)\cite{mehta_book:book}. First introduced in nuclear physics, RMT
has been used with great success in other branches of physics and
mathematics. A notable example was a conjecture by Gorkov and
Eliashberg\cite{Gorkov_Eliashberg_1965} that the single-particle
spectrum of a diffusive metallic grain is controlled by RMT. This
conjecture was proved by Altshuler and
Shklovskii\cite{Altshuler_Shklovskii_86} who used diagrammatic methods
and by Efetov who used the supersymmetry method\cite{Efetov_83}. In
1984 Bohigas, Giannoni and Schmit\cite{bohigas_1:article} conjectured
that RMT could also be employed in the study of ballistic quantum
systems whose dynamics is chaotic in the classical limit. Their
conjecture broadened the area of applicability of RMT enormously and
was supported by numerous ensuing experiments and numerical
simulations \cite{bohigas_1:article, honig:article,
zimmermann:article, deus:article}. The crucial energy scale for the
applicability of RMT is the Thouless energy $E_T=\hbar/\tau_{erg}$,
where $\tau_{erg}$ is the time for a wave packet to spread over the
entire system. For a diffusive system of size $L$, we have $E_T\simeq
\hbar D/L^2$, while for a ballistic/chaotic system we have $E_T\simeq
\hbar v_F/L$, where $v_F$ is the Fermi velocity.

In this paper we consider a system of two quantum dots/nanoparticles
which are coupled by a hopping bridge. The motion of electrons inside
each dot can be either ballistic or diffusive. In the case of ballistic
dots we assume that the dots have irregular shapes leading to
classically chaotic motion, so that RMT is applicable.

RMT Hamiltonians fall into three main
ensembles\cite{mehta_book:book}. These are the Gaussian Orthogonal
Ensemble (GOE), Gaussian Unitary Ensemble (GUE), and Gaussian
Symplectic Ensemble (GSE). They are classified according to their
properties time-reversal (TR). The Hamiltonians invariant with respect
to TR belong to the GOE. An example of GOE is a quantum dot which has
no spin-orbit coupling and is not subject to an external magnetic
field. GUE Hamiltonians, on the contrary, are not invariant with
respect to TR and describe motion in an orbital magnetic field, with
or without spin-orbit coupling.  Hamiltonians from GSE group describe
systems of particles with Kramers degeneracy that are TR invariant but
have no spatial symmetries, and correspond to systems with spin-orbit
coupling but with no orbital magnetic flux. In our paper we only deal
with the first two classes.

For weak magnetic flux the spectral properties of the system deviate
from those predicted by either the GOE or the GUE \cite{sommers:article}.  In such
cases the system is said to be in a crossover\cite{mehta_book:book}. For these systems the
Hamiltonian can be decomposed into real symmetric and real
antisymmetric matrices:

\begin{equation}
  H = \frac{H_S + i X H_A}{\sqrt{1 + X^2}},
\label{eq:11}
\end{equation}
where $X$ is the crossover parameter\cite{aleiner:article} which is
equal, up to factors of order unity, to $\Phi/\Phi_0$, where $\Phi$ is
the magnetic flux through the dot, and $\Phi_0 = h/e$ is the quantum
unit of magnetic flux. Note that the gaussian orthogonal and unitary
ensembles are limiting cases of $X \rightarrow 0$ and $X \rightarrow
1$ respectively.

To understand the meaning of the crossover parameter consider the
Aharonov-Bohm phase shift picked up by a ballistic electron in a
single orbit in the dot:

\begin{equation}
   \Delta\phi = 2\pi \frac{\Phi}{\Phi_0}.
\label{eq:17}
\end{equation}
For one turn the
flux enclosed by the trajectory is proportional to $\Phi = BL^2$,
where $L$ is the size of the dot. After $N$ turns the total flux is
$\Phi_{total} = \sqrt{N}\Phi$, where factor $\sqrt{N}$ originates from
the fact that electron has equal probability to make clockwise or
counterclockwise orbits, and thus does a random walk in the total flux
enclosed. The minimal phase shift for the electron to notice the
presence of the magnetic flux is of the order $2\pi$, and thus the
minimal cumulative flux enclosed by the orbit should be $\Phi_0 =
\sqrt{N} \Phi$. This leads to  $N = (\Phi_0/\Phi)^2$, while the time to make $N$
turns is $\tau = LN/v_f$ (for a ballistic/chaotic dot). From the
Heisenberg uncertainty principle the associated energy scale is:

\begin{equation}
   E_{cross} \approx \frac{\hbar}{\tau} = \frac{E_T}{N} = E_T\left( \frac{\Phi}{\Phi_0} \right)^2,
\label{eq:12}
\end{equation}
where $E_T$ is the ballistic Thouless energy
\cite{altland:article}. For a diffusive dot it should be substituted by
the diffusive Thouless energy $E_T \cong \hbar D/L^2 $. One can see that
when $\Phi$ is equal to $\Phi_0$, $E_X$ is equal to $E_T$ which
means that energy levels are fully crossed over. 

In this paper the reader will encounter many crossover parameters, and
thus many crossover energy scales. By a line of argument similar to
that leading to Eq. \eqref{eq:12}, it can be shown that to every
crossover parameter $X_i$ there is a corresponding energy scale
$E_{X_i} \simeq X_i^2 E_T$.

Breaking the time reversal symmetry of system changes the two-particle
Green's function. While the two-particle Green's function can in
general depend separately on $E_T$, $E_X$, and the measurement
frequency $\omega$, it turns out that in the universal limit $\omega,\
E_X\ll E_T$, it becomes a {\it universal scaling function} of the
ratio $E_X/\omega$. The scaling function describes the modification of 
$\langle G^R(E+\omega)G^A(E)\rangle$ as one moves away from the
``critical'' point $\omega = 0$.
The limits of the scaling function can be understood as follows: If
the measurement frequency $\omega$ is large (small) compared to the
crossover energy scale $E_X$, the $\langle G^R(E+\omega)G^A(E)\rangle$
takes the form of the GOE (GUE) ensemble correlation function.  If
$\omega \sim E_X$, the Green's function describes the system in
crossover regime.

The one-particle Green's function $\langle G^R(E)\rangle$ is not
critical as $\omega \rightarrow 0$, although it gets modified by the
interdot coupling.  The two-particle Green's function $\langle
G^R(E+\omega)G^A(E)\rangle$ always has a diffuson mode
\cite{efetov:book}, that diverges for small $\omega$ in our large-$N$
approximation, which means that our results are valid on scales much
larger than mean level spacing.  This divergence is not physical and
will be cut off by vanishing level correlations for $\omega\ll\delta$
in a more exact calculation\cite{abrikosov:book}.  On the other hand,
the energy scale $\omega$ should be smaller than Thouless energy of
the system for RMT to be applicable. These limitations hold for the
crossover energy $E_X$ as well. In what follows we study the regime
corresponding to $\delta\ll \omega,E_X\le E_T$.

The other term that appears in the two-particle Green's function is a
cooperon mode. In general the cooperon term is gapped if at
least one of the crossover parameters is different from zero. In the case
when the total Hamiltonian of the system is time reversal invariant, all
the crossover parameters are zero and the cooperon, just like the diffuson,
becomes gapless. Finally, when each part of compound system belongs
to the  GUE (the case when all crossover parameters are much larger than
$\omega$) the cooperon term disappears.

Our study has a two-fold motivation. The first part comes from works
on coupled structures with noninteracting particles in acoustic and
electronic systems\cite{waugh:article, weaver:article,
Tschersich_Efetov_00}, and crossovers \cite{falko:article,
french:article, langen:article, pandey:article, sommers:article}.  We
focus on a complete description of the crossover regimes in all three
regions (the two dots and the bridge) and define scaling functions for
the diffuson and cooperon parts of the two-particle Green's
function. Using parameters analogous to $E_X$ we describe crossover
regimes in dots 1 and 2 and the effects of the tunable hopping between
them. Varying these parameters allows us to obtain results for various
physical realizations, when different parts of the compound system
behave as pure GOE, GUE, or belong to the crossover ensemble. In
electronic systems it is easy to break time-reversal by turning on an
external orbital magnetic flux. In acoustic systems one can break
time-reversal by rotating the system or a part thereof. As mentioned
before, the system of two dots coupled by hopping has been
investigated before using supersymmetry
methods\cite{Tschersich_Efetov_00}. However, the authors considered
only the GUE, whereas here we are interested in the full crossover. In
fact, the crossover is essential to the second aspect of our work, as
will become clear immediately.

The second part of our motivation is the possibility of using the
information gained in noninteracting systems to predict the behavior
of interacting systems\cite{adam:article, Adam_Brouwer_Sharma_03,
Alhassid_Rupp_03:condmat, murthy:article}. We consider interacting
systems controlled by the Universal
Hamiltonian\cite{Andreev_Kamenev_81, Brower_Oreg_Halper_99,
Baranger_Ulmo_Glazman_00, Kurland_Aleiner_Altshuler_00}, which is
known to be the interacting low-energy effective
theory\cite{Murthy_Mathur_02:paper, Murthy_Shankar_03,
Murthy_Shankar_Herman_Mathur_04} deep within the Thouless band
$|\varepsilon-\varepsilon_F|\ll E_T$ in the renormalization
group\cite{Shankar_94:paper, Shankar_91} sense for weak-coupling when
the kinetic energy is described by RMT and the Thouless number
$g=E_T/\delta\gg1$. For the GOE the Universal Hamiltonian $H_U$ has
the form\cite{Andreev_Kamenev_81, Brower_Oreg_Halper_99,
Baranger_Ulmo_Glazman_00, Kurland_Aleiner_Altshuler_00}
\begin{equation} 
H_U=\sum\limits_{\a,s}\e_{\a}c^{\dagger}_{\a,s}c_{\a,s}+{U_0\over 2}{\hat N}^2 -J\bS^2+\lambda T^{\dagger} T
\label{univ-ham}
\end{equation}
where ${\hat N}$ is the total particle number, $\bS$ is the total
spin, and $T=\sum c_{\b,\downarrow}c_{\b,\uparrow}$. In addition to
the charging energy, $H_U$ has a Stoner exchange energy $J$ and a
reduced superconducting coupling $\lambda$. This last term is absent in the
GUE, while the exchange term disappears in the GSE.

In this paper we concentrate on the reduced Bardeen-Cooper-Schrieffer
(BCS) coupling $\lambda$ which leads to a mean-field superconducting
state when $\lambda<0$.  Previous work by one of
us\cite{murthy:article} sets the context for our investigation. We
consider an interacting system which has a single-particle symmetry
and a quantum phase transition in the limit $E_T/\delta\to\infty$. An
example relevant to us is a superconducting nanoparticle originally in
the GOE. It has the reduced BCS interaction and time-reversal
symmetry, and the (mean-field) quantum phase transition is between the
normal and superconducting states and occurs at $\lambda=0$. Now
consider the situation when the symmetry is softly broken, so that the
single-particle dynamics is described by a crossover RMT ensemble. It
can be shown \cite{murthy:article} that this step allows us to tune
into the {\it many-body quantum critical
regime}\cite{Chakravarty_Halperin_Nelson_89,
Chakravarty_Halperin_Nelson_88, Sachdev:book} of the interacting
system. Thus, the scaling functions of the noninteracting crossover
are transmuted into scaling functions of the interacting system in the
many-body quantum critical regime. In our example, the orbital
magnetic flux breaks the time-reversal symmetry which is crucial to
superconductivity. When the orbital flux increases to a critical
value, it destroys the mean-field superconducting state. Above the
critical field, or more generically above the critical temperature,
the system is in the quantum critical regime.

\begin{figure*}
  \centerline{
    \mbox{\includegraphics[width=3.5in]{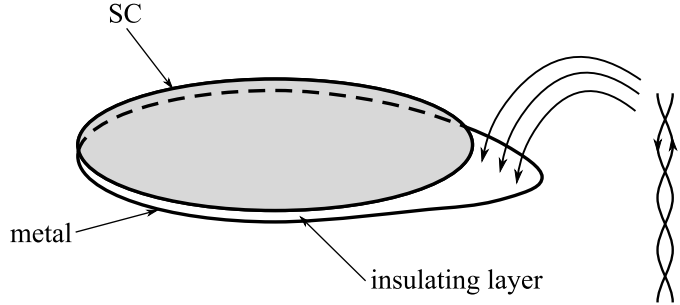}}
  }
  \caption{The system of two vertically coupled quantum dots.}
  \label{vert-coupled}
  \end{figure*}

To be more specific, we consider two vertically coupled quantum dots,
the first of which has an attractive reduced BCS coupling, while the
second has no BCS coupling. Fig. \ref{vert-coupled} shows the
geometry, the reason for which will become clear soon. We apply an
orbital magnetic flux only through (a part of) the second dot, and
observe the effect on the coupled system. Our main results are for the
mean-field critical temperature $T_c$ of the system, and its
magnetization in the normal state (above $T_c$) as a function of the
flux in the normal nanoparticle. Such a system could be realized
physically without too much difficulty, by, for example, growing a
thin film of normal metal (such as $Au$) on an insulating substrate,
then a layer of insulator which could serve as the hopping bridge, and
finally a thin film of superconductor(such as $Al$, which has a
mean-field superconducting transition temperature of around
$2.6K$). The orbital flux can be applied selectively to the $Au$ layer
as shown in Fig. \ref{vert-coupled} by a close pair of oppositely
oriented current carrying wires close to the $Au$ quantum dot, but far
from the $Al$ quantum dot.

The reason for this geometry is that we want to disregard interdot
charging effects entirely and concentrate on the BCS coupling. The
Hamiltonian for the coupled interacting system contains charging
energies for the two dots and an interdot Coulomb
interaction\cite{Adam_Brouwer_Sharma_03}.
\beq
{U_1\over2}N_1^2+{U_2\over2}N_2^2+U_{12}N_1N_2
\eeq
Defining the total number of particles as $N=N_1+N_2$, and the
difference in the number as $n=N_1-N_2$ the interaction can also be written as
\beq
{U_1+U_2+2U_{12}\over16}N^2+{U_1+U_2-2U_{12}\over16}n^2+{U_1-U_2\over4}nN
\eeq
We see that there is an energy cost to transfer an electron from one
dot to the other. This interaction is irrelevant in
the RG sense\cite{Adam_Brouwer_Sharma_03}, but  vanishes only
asymptotically deep within an energy scale defined by the hopping. Our
geometry is chosen so as to make $U_1=U_2=U_{12}$ as nearly as
possible, which can be achieved by making the dots the same thickness
and area, and by making sure that their vertical separation is much
smaller than their lateral linear size. In this case, since $N$ is
constant, we can ignore charging effects entirely. Charging effects
and charge quantization in finite systems can be taken into account
using the formalism developed by Kamenev and
Gefen\cite{kamenev-gefen96}, and futher elaborated by Efetov and
co-workers\cite{efetov-tschersich03,efetov-etal06}.  Since our primary
goal is to investigate quantum critical effects associated with the
BCS pairing interaction, we will assume the abovementioned geometry
and ignore charging effects in what follows.

After including the effect of the BCS interaction, we find the
surprising result that in certain regimes of interparticle hopping
strength, the mean-field transition temperature of the system can {\it
increase} as the flux through the second quantum dot
increases. Indeed, its behavior can be monotonic increasing, monotonic
decreasing, or nonmonotonic as the flux is increased. We can
qualitatively understand these effects by the following
considerations. In the absence of orbital flux, hopping between the
dots reduces $T_c$ since it ``dilutes'' the effect of the attractive
BCS coupling present only in the first dot. The application of an
orbital flux through the second dot has two effects: (i) To raise the
energy of Cooper pairs there, thus tending to localize the pairs in
the first dot and raise the $T_c$. (ii) To cause time-reversal
breaking in the first dot, and reduce $T_c$. The nonmonotonicity of
$T_c$ arises from the competition between these two effects.

Another quantity of interest above the mean-field $T_c$ is the
fluctuation magnetization\cite{Aslamazov_Larkin_68}, which corresponds
to {\it gapped} superconducting pairs forming and responding to the
external orbital flux. In contrast to the case of a single quantum dot
subjected to an orbital flux, we find that the fluctuation
magnetization\cite{Aslamazov_Larkin_68} can be either diamagnetic (the
usual case) or paramagnetic. A paramagnetic magnetization results from
a free energy which decreases as the flux increases. The origin of
this effect is the interplay between the localizing effect of high
temperature or the orbital flux in the second dot on the one hand, and
the reduced BCS interaction on the other.

The regimes we describe should be distinguished from other
superconducting single-particle RMT ensembles discovered in the past
decade\cite{altland-zirnbauer1,altland-zirnbauer2}, which apply to a
normal mesoscopic system in contact with two superconductors with a
phase difference of $\pi$ between their order
parameters\cite{altland-zirnbauer1} (so that there is no gap in the
mesoscopic system despite Andreev reflection), or to a mesoscopic
$d$-wave superconducting system\cite{altland-zirnbauer2}. In our case,
the symmetry of the superconducting interaction is $s$-wave. However,
the most important difference is that we focus on quantum critical
fluctuations, which are inherently many-body, while the RMT classes
described previously are single-particle
ensembles\cite{altland-zirnbauer1,altland-zirnbauer2}.

This paper is organized as follows. In Sec. \ref{single_dot} we review
the basic steps of calculating the one-particle and two-particle
Green's functions for a single dot. Then in Sec. \ref{double_dot} we
present the system of Dyson equations for the one-particle Green's
function in the case of two coupled dots and solve it in the limit of
weak coupling.  In addition, we set up and solve the system of four
Bethe-Salpeter equations for the two-particle Green's function.  In
Sec. \ref{2_quantum dots} we apply our results to the system of
superconducting quantum dot weakly coupled to other quantum dot made
from a normal metal. We end with our conclusions, some caveats, and
future directions in Sec. \ref{conclusion}.
%
%
%
      \section{Review of results for a single dot.}\label{single_dot}
%
%
  Our goal in this section is to calculate the statistics of one and
  two-particle Green's functions for an uncoupled dot in a GOE$\to$
  GUE crossover (see appendix \ref{apnx:A}, and
  \cite{aleiner:article} for more details), starting from the series expansion of
  Green's function:

\begin{equation}
  \langle\beta|G^R(E)|\alpha\rangle = G^R_{\alpha\beta}(E) = \left( \frac{1}{E^{+} - H} \right)_{\alpha\beta} = 
                          \frac{\delta_{\alpha\beta}}{E^+}+\frac{H_{\alpha\beta}}{(E^+)^2}+
                          \frac{H^2_{\alpha\beta}}{(E^+)^3}+\ldots.
\end{equation}
We are interested in averaging this expansion over the appropriate
random matrix ensemble. The corresponding Dyson equation for averaged
Green's function is:
\begin{equation}
        \includegraphics{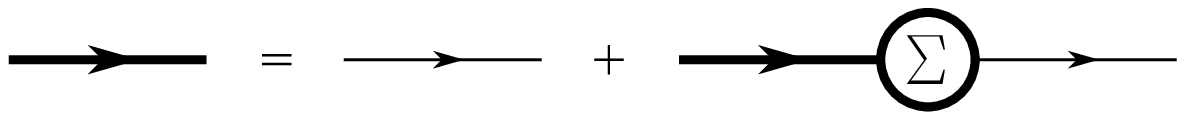}
\label{eq:13}
\end{equation}
 The bold line denotes the averaged propagator $\langle G^R(E)
\rangle$ and regular solid line defines the bare propagator $1/E^+$
with $E^+ = E + i\eta$, where $\eta$ is infinitely small positive
number. Here $\Sigma$ stands for self-energy and is a sum of all
topologically different diagrams.

One can solve Dyson equation approximating self-energy only by first leading term and find:
\begin{equation}
  \Sigma = \frac{E}{2} - \frac{i}{2}\sqrt{\left( \frac{2N\delta}{\pi} \right)^2 - E^2},
\end{equation}
  where $\delta$ is the mean level spacing. This approximation works
  only for $E \gg \delta$. As $E$ gets comparable with $\delta$, other
  terms in expansion for $\Sigma$ should be taken into account.

  Then, the average of the one-particle Green's function is given by:
\begin{equation}
  \langle G^R_{\alpha\beta}(E)\rangle = \frac{\delta_{\alpha\beta}}
                                {\frac{E}{2} + \frac{i}{2}\sqrt{\left( \frac{2N\delta}{\pi} \right)^2 - E^2}}
\end{equation}
  Next, we repeat the procedure for the averaged two-particle Green's
  function, which can be represented by the series:
\begin{equation}
        \includegraphics{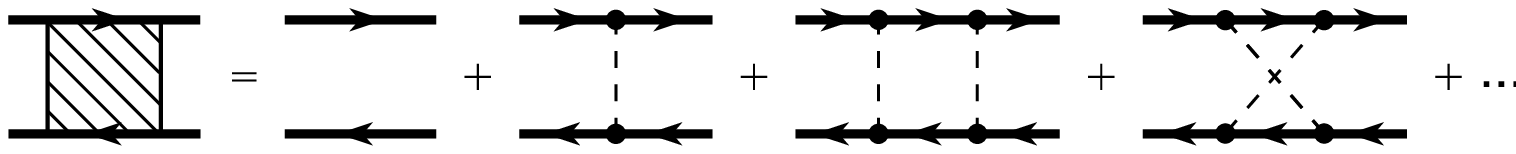}
\end{equation}
   where two bold lines on the left hand side denote $\langle G^R(E +
\omega)G^A(E)\rangle$. The leading contribution comes from ladder and
maximally crossed diagrams. The sum of these diagrams can be
conveniently represented by Bethe-Salpeter equation. For example, the
contribution of all the ladder diagrams can be expressed in closed
form by:
\begin{equation}
        \includegraphics{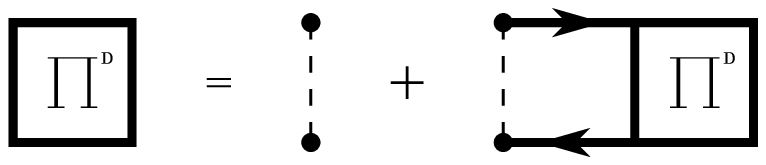}
\end{equation}
   where $\Pi^D$ is a self-energy. For maximally crossed diagrams we have similar equation:

\begin{equation}
        \includegraphics{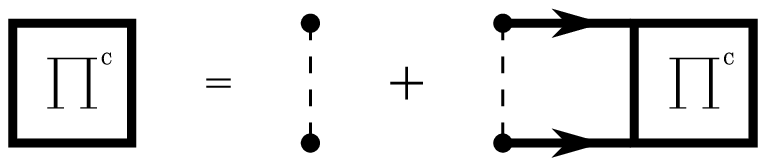}
\end{equation}
   where $\Pi^D$ and $\Pi^C$ are related to the connected part of two-particle Green's function as:

\begin{equation}
   \includegraphics{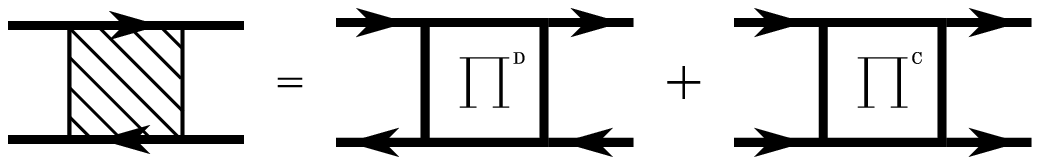}
\end{equation}

  In the limit of $\omega$ being much smaller than bandwidth ($\omega
\ll N\delta$), the two-particle Green's function (connected part) is
expressed as:
\begin{equation}
  \langle G^R_{\alpha\gamma}(E + \omega)G^A_{\delta\beta}(E) \rangle = \frac{2\pi}{N^2\delta}
                             \frac{\delta_{\alpha\beta} \delta_{\gamma\delta}}{-i\omega} + 
                                                         \frac{2\pi}{N^2\delta}
                             \frac{\delta_{\alpha\delta} \delta_{\gamma\beta}}{-i\omega}
                                 \frac{1}{1 + i\frac{E_X}{\omega}}
\end{equation}
  The second term is a contribution of maximally crossed diagrams. $E_X$ is a crossover energy scale, 
  connected to the crossover parameter as $E_X = 4X^2N\delta/\pi$.

Depending on values of $E_X$ one can speak of different types of
averaging. If $E_X \ll \omega$, we get average over GOE ensemble, if
$E_X$ is of order $\omega$, averaging is performed over ensemble being
in crossover, and, if $E_X \gg \omega$, contribution of maximally
crossed diagrams can be disregarded, thus going to the limit of the
GUE ensemble.
\section{Two coupled dots.}\label{double_dot}

  Next we discuss general framework of our calculation and calculate correlation functions for 
our system of interest, which is two weakly coupled quantum dots (see appendix \ref{apnx:B} for more technical details).
 The Hamiltonian for this system can be represented as:
\begin{equation}
  H_{tot} = \begin{pmatrix}
               H_{1} &  0      \\
               0     &  H_{2}
            \end{pmatrix}
          +
            \begin{pmatrix}
               0     &  V      \\
               V^{\dagger} &  0
            \end{pmatrix}
          =
            \begin{pmatrix}
               H_{1} & V \\
               V^{\dagger} & H_{2}
            \end{pmatrix}.
\end{equation}
where $H_{1}$ and $H_{2}$ are the Hamiltonians of uncoupled dots $1$ and $2$. The coupling is realized
by a matrix $V$. The elements of $H_1$, $H_{2}$, and $V$ are statistically independent random variables. 
We assume that both dots and the hopping bridge are in crossover regimes, characterized by parameters 
$X_1$, $X_2$, and $\Gamma$ respectively.

In the crossover matrices $H_i$ and $V$ are given by:
\begin{equation}
  H_i = \frac{ H^S_i + iX_i H^A_i }{\sqrt{1 + X_i^2}}, \ i=1,2; \hspace{1cm}
  V = \frac{ V^R + i\Gamma V^I }{\sqrt{1 + \Gamma^2}},
\end{equation}
  where $H^{S,A}_i$ is a symmetric (antisymmetric) part of $H_i$, and $V^{R,I}$ is real (imaginary) matrix.
  In what follows we assume that the bandwidths in dot $1$ and dot $2$ are the same. That is, 
  $N_1\delta_1 = N_2\delta_2$. This should not make any difference in the universal limit 
  $N \rightarrow \infty$. In addition we introduce the parameter $\xi$ -- the ratio of mean level spacing in two dots:
  $\xi = \delta_1/\delta_2$. For each realization of matrix elements of the Hamiltonian $H_{tot}$, the 
 Green's function of this system can be computed as follows:
\begin{equation}
  G = (I \otimes E - H)^{-1} = \begin{pmatrix}
                                   E-H_1 & -V \\
                                   -V^{\dagger} & E-H_{2}
                               \end{pmatrix}^{-1}
                             = \begin{pmatrix}
                                   G_{11} & G_{12} \\
                                   G_{21} & G_{22}
                               \end{pmatrix}.
\end{equation}

  Each element of $G$ has the meaning of a specific Green's function.
For example, $G_{11}$ and $G_{22}$ are the Green's functions that describe particle propagation 
in dots $1$ and $2$ respectively. On the other hand, $G_{12}$ and $G_{21}$ are the Green's
functions representing travel from one dot to another.

Calculating $(I\otimes E - H)^{-1}$ one finds the components of $G$. For example,

\begin{equation}
  G_{11} = \left[ (E-H_1) - V(E-H_2)^{-1}V^{\dagger} \right]^{-1} = G_1
                   + G_1 V G_2 V^{\dagger} G_1
                   + G_1 V G_2 V^{\dagger} G_1 V G_2 V^{\dagger} G_1 + \ldots
\label{eq:2}
\end{equation}
  where $G_1$ and $G_2$ are bare propagators in dot 1 and dot 2
defined by $G_1 = (E - H_1)^{-1}$ and $G_2 = (E - H_2)^{-1}$.

To find the ensemble average of $G_{11}$ one needs to average the whole expansion \eqref{eq:2} 
term by term.
For coupled dots $G_{ij}$ interrelated and in large N approximation can be found from 
the following system of equations:

\begin{equation}
   \includegraphics{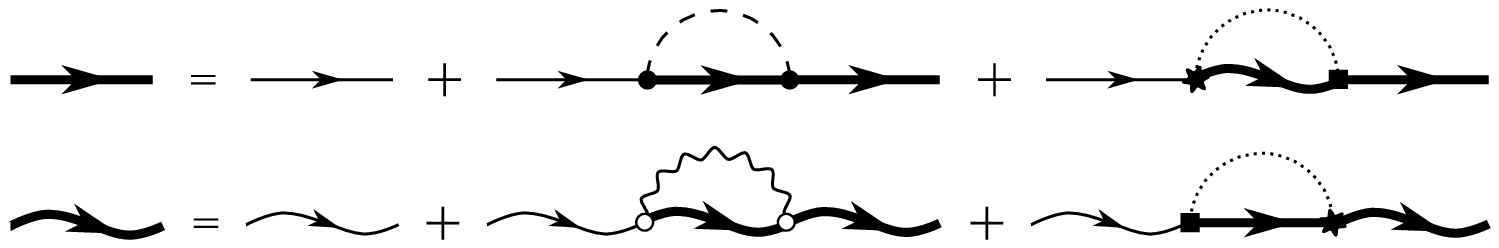}
\label{eq:3}
\end{equation}

The bold straight and wavy lines with arrows represent averaged Green's functions 
$\langle G_{\alpha\beta,1}(E)\rangle$ and $\langle G_{ij,2}(E)\rangle$ respectively, while 
regular solid lines are bare propagators in dots 1 and 2. The dotted line describes pairing 
between hopping matrix elements $V$, and the dashed (wavy) line denotes pairing between matrix 
elements of $H_{1}$ ($H_{2}$).

The system \eqref{eq:3} accounts for all possible diagrams without
line crossing. Diagrams containing crossed lines of any type are
higher order in $1/N$ and can be neglected when $N \rightarrow
\infty$. If the hopping between dots is zero, this system decouples
into two separate Dyson equations for each dot. In the case of weak
coupling ($U \ll 1$), where $U$ is a parameter controlling the
strength of coupling between dots, this system can be readily solved.
As zero approximation, we use results for a single dot.

In this approximation one-particle Green's function for dot 1 and dot 2 are calculated as follows:

\begin{equation}
 \begin{split}
  \langle G^R_{\alpha\beta,1}(E)\rangle &= \frac{\langle G^R_{\alpha\beta,0}(E)\rangle}{1-U\sqrt{\xi}\frac{\Sigma_0}{E-2\Sigma_0} }
                           = \frac{\delta_{\alpha\beta}}
                                          {\left( \frac{N_1\delta_1}{\pi} \right) 
                                           \left[ \epsilon + i\sqrt{1-\epsilon^2} \right]}
                             \frac{1}{ \left[ 1+\frac{U\sqrt{\xi}}{2} 
                             \left( 1+i\frac{\epsilon}{\sqrt{1-\epsilon^2}} \right) \right] } \\
  \langle G^R_{ij,2}(E)\rangle &= \frac{\langle G^R_{ij,0}(E)\rangle}{1-\frac{U}{\sqrt{\xi}}\frac{\Sigma_0}{E-2\Sigma_0} }
                           = \frac{\delta_{ij}}
                                  {\left( \frac{N_2\delta_2}{\pi} \right)
                                   \left[ \epsilon + i\sqrt{1-\epsilon^2} \right]} 
                             \frac{1}{\left[ 1+\frac{U}{2\sqrt{\xi}}
                             \left( 1+i\frac{\epsilon}{\sqrt{1-\epsilon^2}} \right) \right] },
 \end{split}
\end{equation}
  where $\epsilon$ is a dimensionless energy $\epsilon = \pi E/2N\delta$. We used subindex $0$ in 
$\Sigma_0$ and $\langle G^R_0(E) \rangle$ to denote solutions for one uncoupled dot.

In the large ${\it N}$ approximation the contribution to the two-particle
Green's function comes from ladder diagrams and maximally crossed
diagrams. It is convenient to sum them separately. The ladder diagram
contribution can be found from the following system of equations:

\begin{equation}
  \includegraphics{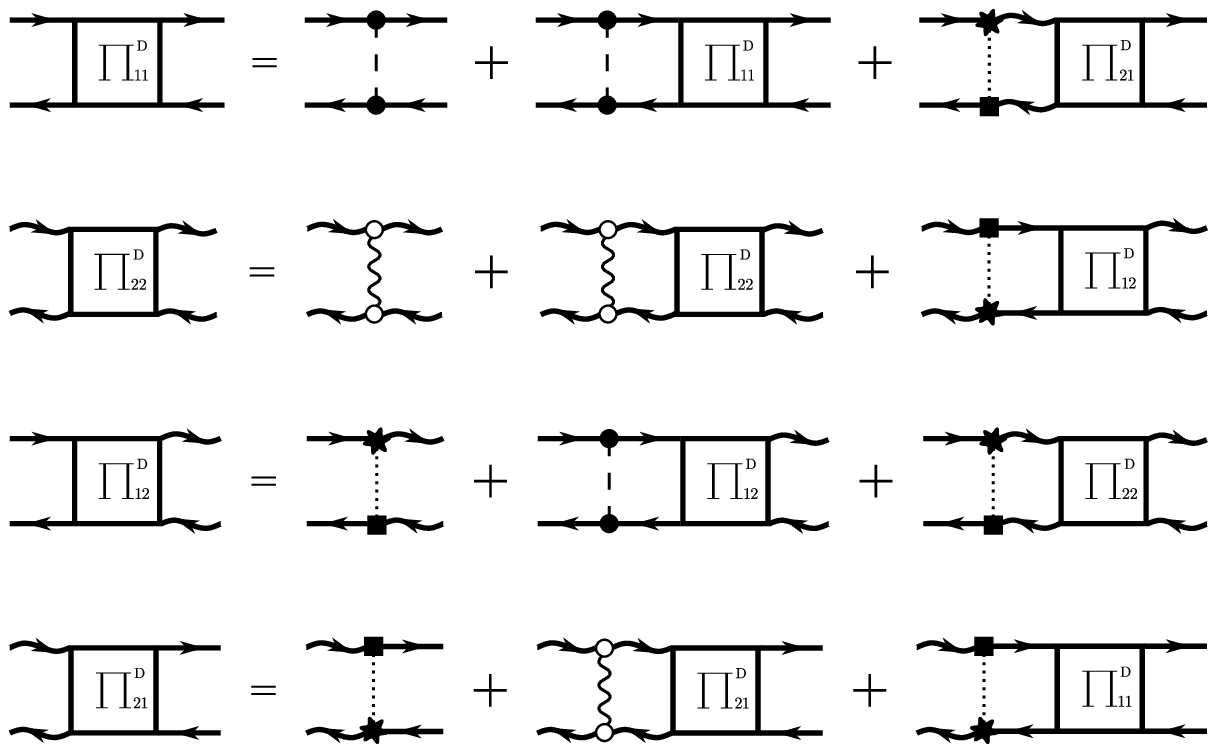}
\label{eq:6}
\end{equation}
  where $\Pi^D_{ij}$ with proper external lines denote various
two-particle Green's functions. As in the case of the one-particle
Green's function equations, if the inter-dot coupling is zero, the
system reduces to two Bethe-Salpeter equations for uncoupled dots.

The system of four equations \eqref{eq:6}  can be broken into two systems of two
equations to get:

\begin{equation*}
 \begin{split}
  \langle G^R_{\alpha\gamma,1}(E+\omega) G^A_{\delta\beta,1}(E)\rangle_{D1} &= 
                   \frac{2\pi}{N_1^2\delta_1} \frac{\delta_{\alpha\beta}\delta_{\gamma\delta}}{-i\omega}
                    g_{D1} \\
  \langle G^R_{il,2}(E+\omega) G^A_{kj,2}(E)\rangle_{D2} &=
                   \frac{2\pi}{N_2^2\delta_2} \frac{\delta_{ij}\delta_{lk}}{-i\omega}
                    g_{D2},
 \end{split}
\end{equation*}
   where $g_{D}$ are the scaling functions of diffusion terms in dot 1 and dot 2 defined by:
\begin{equation}
 \begin{split}
      g_{D1} &= \frac{1+\frac{i}{\sqrt{\xi}} \frac{E_U}{\omega}}
                        {1+i(\sqrt{\xi} + \frac{1}{\sqrt{\xi}}) \frac{E_U}{\omega}} \\
      g_{D2} &= \frac{1+i\sqrt{\xi} \frac{E_U}{\omega}}
                        {1+i(\sqrt{\xi} + \frac{1}{\sqrt{\xi}}) \frac{E_U}{\omega}}.
 \end{split}
\end{equation}
 Here $E_U = 2UN\delta/\pi$ is the interdot coupling energy
scale. These dimensionless functions show how diffusion part is
modified due to the coupling to another dot.

  Next, for the maximally crossed diagrams the system of equations we have:

\begin{equation}
  \includegraphics{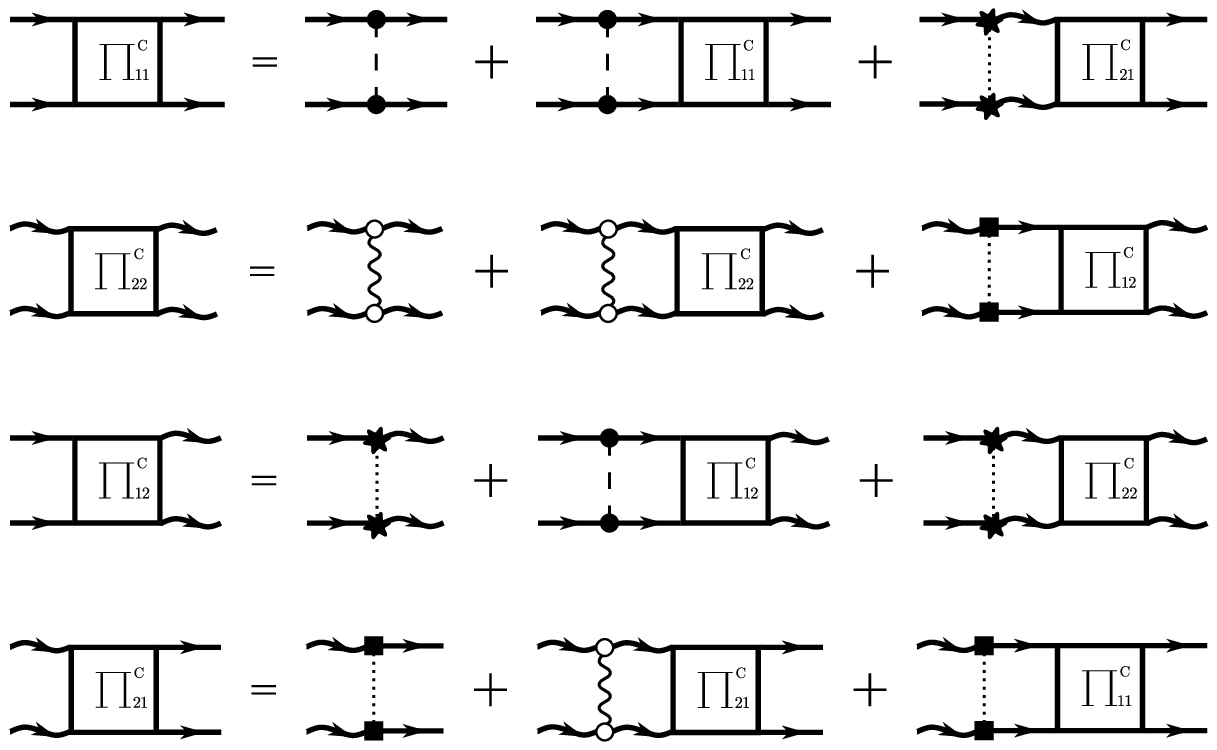}
\label{eq:9}
\end{equation}

  The subsequent solution of this system produces:

\begin{equation}
  \begin{split}
  \langle G^R_{\alpha\gamma,1}(E+\omega) G^A_{\delta\beta,1}(E)\rangle_{C1} &=
         \frac{2\pi}{N_1^2\delta_1} \frac{\delta_{\alpha\delta}\delta_{\gamma\beta}}{-i\omega}
         g_{C1} \\
  \langle G^R_{il,2}(E+\omega) G^A_{kj,2}(E)\rangle_{C2} &=
         \frac{2\pi}{N_2^2\delta_2} \frac{\delta_{ik}\delta_{lj}}{-i\omega}
         g_{C2},
  \end{split}
\end{equation}
   where $g_{C}$ are the scaling functions for cooperon term defined according to:

\begin{equation}
  \begin{split}
  g_{C1} &=
                   \frac{1+\frac{i}{\sqrt{\xi}} \frac{E_U}{\omega} + i\frac{E_{X_2}}{\omega} }
                        {1+i\frac{E_{X_1}+E_{X_2}}{\omega} - \frac{E_{X_1} E_{X_2}}{\omega^2} -
                         \frac{E_{X_1}E_U}{\sqrt{\xi}\omega^2} - \frac{\sqrt{\xi}E_{X_2}E_U}{\omega^2} +
                         i\left( \sqrt{\xi} + \frac{1}{\sqrt{\xi}} \right) \frac{E_U}{\omega}
                         \left( 1+i\frac{E_{\Gamma}}{\omega} \right) } \\
  g_{C2} &=
                   \frac{1 + i\sqrt{\xi} \frac{E_U}{\omega} + i\frac{E_{X_1}}{\omega} }
                        {1+i\frac{E_{X_1}+E_{X_2}}{\omega} - \frac{E_{X_1} E_{X_2}}{\omega^2} -
                         \frac{E_{X_1}E_U}{\sqrt{\xi}\omega^2} - \frac{\sqrt{\xi}E_{X_2}E_U}{\omega^2} +
                         i\left( \sqrt{\xi} + \frac{1}{\sqrt{\xi}} \right) \frac{E_U}{\omega}
                         \left( 1+i\frac{E_{\Gamma}}{\omega} \right) }.
  \end{split}
\end{equation}

  Here $E_{X_{1,2}} = 4X_{1,2}^2 N\delta/\pi$, and $E_{\Gamma} =
4\Gamma^2 E_U/(\sqrt{\xi}+\frac{1}{\sqrt{\xi}})$ are the crossover
energy scales, describing transition from GOE to GUE ensemble in dot 1
and dot 2, as well as in hopping bridge $V$.

  As we determined how the scaling function $g_{C}$ modifies cooperon part of two-particle
 Green's function and depends on the crossover energy scales defined
 above, we are ready to proceed with write up the connected part of the total two-particle 
Green's function, which is a  sum of diffuson and cooperon parts:

\begin{equation}
  \begin{split}
  \langle & G^R_{\alpha\gamma,1}(E+\omega) G^A_{\delta\beta,1}(E)\rangle =
             \frac{2\pi}{N_1^2\delta_1} \frac{\delta_{\alpha\beta}\delta_{\gamma\delta}}{-i\omega}
             \frac{1+\frac{i}{\sqrt{\xi}} \frac{E_U}{\omega}}
                  {1+i(\sqrt{\xi} + \frac{1}{\sqrt{\xi}}) \frac{E_U}{\omega}} + \\
                &  \frac{2\pi}{N_1^2\delta_1} \frac{\delta_{\alpha\delta}\delta_{\gamma\beta}}{-i\omega}
                   \frac{1+\frac{i}{\sqrt{\xi}} \frac{E_U}{\omega} + i\frac{E_{X_2}}{\omega} }
                        {1+i\frac{E_{X_1}+E_{X_2}}{\omega} - \frac{E_{X_1} E_{X_2}}{\omega^2} -
                         \frac{E_{X_1}E_U}{\sqrt{\xi}\omega^2} - \frac{\sqrt{\xi}E_{X_2}E_U}{\omega^2} +
                         i\left( \sqrt{\xi} + \frac{1}{\sqrt{\xi}} \right) \frac{E_U}{\omega}
                         \left( 1+i\frac{E_{\Gamma}}{\omega} \right) }
  \end{split}
\label{eq:18}
\end{equation}
\begin{equation}
  \begin{split}
  \langle & G^R_{il,2}(E+\omega) G^A_{kj,2}(E)\rangle = 
          \frac{2\pi}{N_2^2\delta_2} \frac{\delta_{ij}\delta_{lk}}{-i\omega}
          \frac{1+i\sqrt{\xi} \frac{E_U}{\omega}}
               {1+i(\sqrt{\xi} + \frac{1}{\sqrt{\xi}}) \frac{E_U}{\omega}} + \\
                &  \frac{2\pi}{N_2^2\delta_2} \frac{\delta_{ik}\delta_{lj}}{-i\omega}
                   \frac{1 + i\sqrt{\xi} \frac{E_U}{\omega} + i\frac{E_{X_1}}{\omega} }
                        {1+i\frac{E_{X_1}+E_{X_2}}{\omega} - \frac{E_{X_1} E_{X_2}}{\omega^2} -
                         \frac{E_{X_1}E_U}{\sqrt{\xi}\omega^2} - \frac{\sqrt{\xi}E_{X_2}E_U}{\omega^2} +
                         i\left( \sqrt{\xi} + \frac{1}{\sqrt{\xi}} \right) \frac{E_U}{\omega}
                         \left( 1+i\frac{E_{\Gamma}}{\omega} \right) }
  \end{split}
\end{equation}

In general, the coupling between dots changes the bandwidth of each dot. Corrections to the bandwidth
are of the order of $U$ and can be neglected for weak coupling. Calculating approximations to the
second order in $U$ one can be ensure that one-particle and two-particle Green's functions can be treated
perturbatively.

  \begin{figure*}
  \centerline{
    \mbox{\includegraphics[width=3.5in]{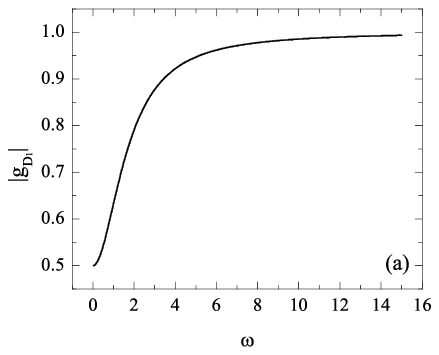}}
    \mbox{\includegraphics[width=3.5in]{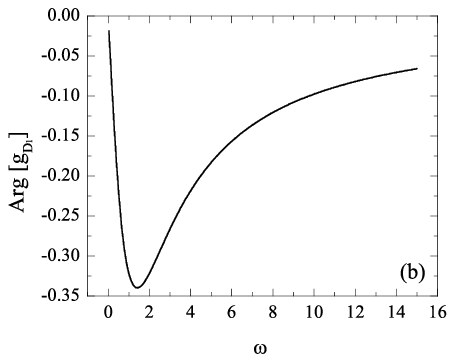}}
  }
  \centerline{
    \mbox{\includegraphics[width=3.5in]{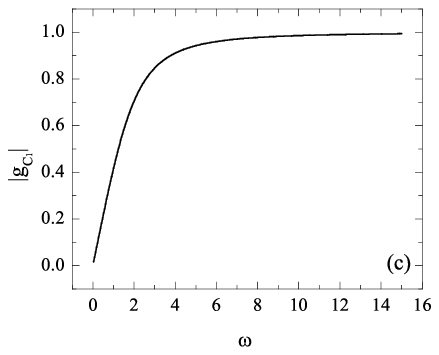}}
    \mbox{\includegraphics[width=3.5in]{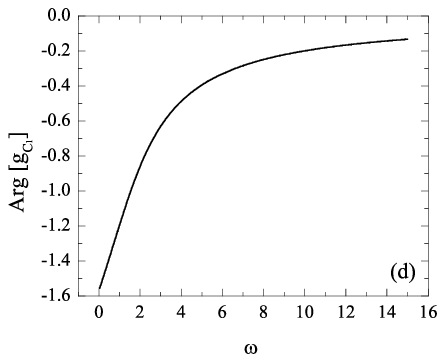}}
  }
  \caption{Absolute value and phase of diffuson (a,b) and cooperon (c,d) scaling functions in dot 1.
            Frequency $\omega$ is measured in units of $E_U$. For these graphs the crossover 
            parameters are: $E_{X_1}/E_U = E_{X_2}/E_U=1$, $E_{\Gamma}/E_U=0.8$, $\xi=1$.}
  \label{View1}
  \end{figure*}

   Diagrams on Fig.\ref{View1} show the typical behavior of absolute value and phase of scaling functions $g_{D}$ 
and $g_{C}$ in dot 1. All energy parameters are measured in units of $E_U$.

%
%

  Next we analyze the temporal behavior of the computed statistical
  characteristics.  The Fourier transform of the two-particle Green's
  function shows the time evolution of the density matrix of the
  system. One can observe that the diffuson part of $\langle G^RG^A
  \rangle$ diverges for small $\omega$. To get the correct behavior we
  replace $1/\omega$ with $\omega/(\omega^2+\eta^2)$, and take $\eta$
  to zero in the final result. As for the cooperon term, it stays
  regular in the small $\omega$ limit if at least one of the crossover
  parameters differs from zero.

First of all, we look at the Fourier transform of $\langle G^RG^A \rangle$ in the first dot. We have

\begin{multline}
  \langle G^R_{\alpha\gamma,1}(t) G^A_{\delta\beta,1}(t)\rangle =
        \delta_{\alpha\beta} \delta_{\gamma\delta} \frac{2}{N_1(1+\xi)}
    \left[ \frac{1}{2} + \xi e^{-(\sqrt{\xi} +
           \frac{1}{\sqrt{\xi}}) E_U t} \right] \\
  + \delta_{\alpha\delta} \delta_{\gamma\beta} \frac{2}{N_1}
       \left[
           1 + \frac{ E_{X_2} + \frac{E_U}{\sqrt{\xi}} }{a_+ - a_-}
           \left( e^{-t a_{-}} - e^{-t a_{+}} \right)
       \right],
\end{multline}
  where $a_{\pm}$ depend on the crossover parameters (see
  Eq. \eqref{eq:10} in appendix \ref{apnx:C})

  Then, for the corresponding quantity in the second dot the Fourier transform produces:

\begin{multline}
  \langle G^R_{il,2}(t) G^A_{kj,2}(t)\rangle =
           \delta_{ij} \delta_{lk} \frac{2\xi}{N_2(1+\xi)}
  \left[
     \frac{1}{2} + \frac{1}{\xi} e^{-(\sqrt{\xi} +
               \frac{1}{\sqrt{\xi}}) E_U t} \right] \\
    + \delta_{ik} \delta_{lj}\frac{2}{N_2}
       \left[
           1 + \frac{ E_{X_1} + \sqrt{\xi}E_U }{a_+ - a_-}
           \left( e^{-t a_{-}} - e^{-t a_{+}} \right)
       \right].
\end{multline}

\section{Two coupled metallic quantum dots}\label{2_quantum dots}

  In this section we apply the results obtained in the previous
sections to an interacting system. We consider two vertically coupled
metallic quantum dots, as shown in Fig. \ref{vert-coupled}, the first of which is
superconducting and the second noninteracting. For simplicity the
quantum dots are assumed to have the same level spacing ($\xi =
1$). The calculations presented in this section can be extended to the
case $\xi\ne 1$ in a straightforward way. The first (superconducting)
quantum dot and the hopping bridge belong to the GOE ensemble. A
nonzero orbital magnetic flux penetrating the second (noninteracting)
quantum dot drives it into the GOE to GUE crossover described by the
crossover energy scale $E_{X_2}$. The other crossover energy scale
$E_U$ describes the hopping between the quantum dots. Because of this
hopping one can observe a nonzero magnetization in the first particle
caused by a magnetic flux through the second particle. Roughly
speaking, when the electrons in the first dot travel to the second and
return they bring back information about the orbital flux.

We wish to compute the magnetization as a function of orbital flux, as
well as the mean-field critical temperature. It should be noted that
since the quantum dot is a finite system, there cannot be any true
spontaneous symmetry breaking. However, when the mean-field
superconducting gap $\Delta_{BCS}\gg\delta$, the mean-field
description is a very good
one\cite{Schechter_Oreg_Imry_Levinson_03,Ambegaokar_Eckern_90,
Ambegaokar_Eckern_99_epl}. Recent numerical calculations have
investigated the regime $\Delta_{BCS}\simeq\delta$ where quantum
fluctuations are strong\cite{alhassid-fang-schmidt06}. We will focus
on the quantum critical regime of the system above the mean-field
critical temperature/field, so we do not have to worry about
symmetry-breaking.
 
We start with BCS crossover Hamiltonian for the double-dot system
including the interactions in the first dot and the hopping between
the dots\cite{murthy:article}:

\beqr
  H_{BCSX_{2}} =& \sum\limits_{\mu_0\nu_0}H^{(1)}_{\mu_0\nu_0}c^{\dagger}_{\mu_0s}c_{\nu_0s}-\lambda T^{\dagger}T+\nonumber\\
&\sum\limits_{i_0j_0s}H^{(2)}_{i_0j_0s}c^{\dagger}_{i_0s}c_{i_0s}+\sum\limits_{\mu_0i_0}V_{\mu_0i_0}(c^{\dagger}_{\mu_0s}c_{i_0s}+h.c.)\nonumber\\
=& \sum\limits_{\mu s} \epsilon_{\mu} c^{\dagger}_{\mu,s} c_{\mu,s} 
                                              - \delta \tilde{\lambda} T^{\dagger}T,
\label{hbcsx2}
\eeqr
where $H^{(2)}$ contains the effect of the orbital flux through the
second quantum dot. Here $T,\ T^{\dagger}$ are the operators which
appear in the Universal Hamiltonian, and are most simply expressed in
terms of electron creation/annihilation operators in the original GOE
basis of the first dot (which we call $\mu_0,\nu_0$) as
\begin{equation}
T=\sum_{\mu_0}c_{\mu_0,\downarrow} c_{\mu_0,\uparrow}
\end{equation}
Now we need to express the operators $c_{\mu_0,s}$ in terms of the
eigenoperators of the combined single-particle Hamiltonian of the
system of two coupled dots. The result is 
\begin{equation}
  T = \sum_{\mu\nu} M_{\mu\nu} c_{\nu,\downarrow} c_{\mu,\uparrow}, \hspace{1cm} 
  M_{\mu\nu} = \sum_{\mu_0} \psi_{\mu}(\mu_0) \psi_{\nu}(\mu_0),
\end{equation}
  where $\epsilon_{\mu}$ denotes the eigenvalues of the total system,
$c_{\mu,s}$ operator annihilates electron in the orbital state $\mu$
with spin $s$, $\psi_{\mu}(\mu_0)$ is the eigenvector of the compound
system, $\delta$ is the mean level spacing of a single isolated dot,
$\tilde{\lambda}>0$ is the attractive dimensionless BCS coupling valid
in region of width $2\omega_D$ around the Fermi energy. Note that while
the indices $\mu,\nu$ enumerate the states of the total system, the
index $\mu_0$ goes only over the states of the first dot, since the
superconducting interaction is present only in the first dot.

To study the magnetization of the first quantum dot in the crossover
we follow previous work by one of us\cite{murthy:article}: We start
with the partition function $Z=Tr(exp{-\b H})$ where $\b=1/T$ is the
inverse temperature. We convert the partition function into an
imaginary time path integral and use the Hubbard-Stratanovich identity
to decompose the interaction, leading to the imaginary time Lagrangian
\beq 
{\cal L}={|\s|^2\over\delta{\tilde{\lambda}}}-\sum\limits_{\mu,s}{\bar{c}}_{\mu,s}(\partial_{\tau}-\epsilon_{\mu})c_{\mu,s}+\s{\bar T}+\bss T
\eeq 
where $\s,\bss$ are the bosonic Hubbard-Stratanovich fields
representing the BCS order parameter and ${\bar{c}},c$ are Grassman
fields representing fermions.  The fermions are integrated out, and as
long as the system does not have a mean-field BCS gap, the resulting
action for $\s,\bss$ can be expanded to second order to obtain 
\beqr
S_{eff}\approx {\delta\over\b}\sum\limits_{n}& |\s(i\omega_n)|^2
({1\over{\tilde{\lambda}}}-f_n(\b,{E_{X}},\omega_D)) \\
f_n(\b,{E_{X}},\omega_D)=&\delta\sum\limits_{\mu\nu} |M_{\mu\nu}|^2
{1-N_F(\epsilon_{\mu})-N_F(\epsilon_{\nu})\over
\epsilon_{\mu}+\epsilon_{\nu}-i\omega_n} 
\eeqr
where $\omega_n=2\pi n/\b$, and the sums are restricted to
$|\epsilon_{\mu}|,|\epsilon_{\nu}|<\hbar\omega_D$. We see that the
correlations between different states $\mu,\nu$ play an important
role. Deep in the crossover (for ${E_{X}}\gg\delta$) we can replace
$|M_{\mu\nu}|^2$ by its ensemble average\cite{murthy:article}.  We
will also henceforth replace the summations over energy eigenstates by
energy integrations with the appropriate cutoffs. In previous
work\cite{murthy:article} the statistics\cite{adam:article,
Adam_Brouwer_Sharma_03, Alhassid_Rupp_03:condmat} of $|M_{\mu\nu}|^2$
was used to obtain analytical results for this expression.

  The (interacting part of the) free energy of the system in the quantum critical regime is
  given by \cite{murthy:article}:

\begin{equation}
  \beta F = \sum_{n} \ln(1 - \tilde{\lambda} f(i\omega_n,\beta,E_{X_2})),
\end{equation}
  where $f$ is the scaling function given by 
expression:

\begin{equation}
  f(i\omega_n,\beta,E_{X_2}) = \delta\sum_{\mu\nu} \lvert M_{\mu\nu} \rvert^2 
                            \frac{1- n_{\mu}(\beta) - n_{\nu}(\beta)}
                                 {\epsilon_{\mu} + \epsilon_{\nu} - i\omega_n},
\end{equation}
  $n_{\nu}(\beta) = (1 + \exp(\beta\epsilon_{\nu}))^{-1}$ is the
  Fermi-Dirac distribution. We have shifted the energy so that the
  chemical potential is 0.

Converting this double sum into integral and substituting $\lvert M_{\mu\nu} \rvert^2$ by its 
ensemble average (see Appendices \ref{apnx:D} and \ref{apnx:E}), we get:

\begin{equation}
  f_n =  \frac{E_U}{\pi} \int^{\omega_D}_{-\omega_D} d\epsilon_1 d\epsilon_2 
          \frac{(\epsilon_1 - \epsilon_2)^2 + E_{X_2}E_U + E_{X_2}^2}
               {((\epsilon_1 - \epsilon_2)^2 - E_{X_2}E_U )^2 + (E_{X_2} + 2E_U)^2 (\epsilon_1 - \epsilon_2)^2}
           \frac{\tanh(\frac{\beta\epsilon_1}{2}) + \tanh(\frac{\beta\epsilon_2}{2}) }
                {\epsilon_1 + \epsilon_2 - i\omega_n },
\end{equation}
  where $\omega_D$ is the Debye frequency, and $\beta  = 1/k_{B}T$ is the inverse temperature.

 One can decompose the ratio in the first part of integrand into two Lorentzians
 to get \cite{murthy:article}:

\begin{multline}
 f_n =  \frac{E_U}{2E_1} \frac{E_{X_2}^2 + E_U E_{X_2} - E_1^2}{E_2^2 - E_1^2} 
         \ln \left[ \frac{4(\hbar\omega_D)^2 + \omega_n^2 }{C'/\beta^2 + (E_1+|\omega_n|)^2 } \right]\\
       + \frac{E_U}{2E_2} \frac{E_2^2 - E_{X_2}^2 - E_U E_{X_2} }{E_2^2 - E_1^2}
         \ln \left[ \frac{4(\hbar\omega_D)^2 + \omega_n^2 }{C'/\beta^2 + (E_2+|\omega_n|)^2 } \right].
\label{scalingchi}\end{multline} 
  Here $C' \approx 3.08$ and $E_{1,2}$ depend on crossover energy scales as follows:

\begin{equation}
  E^2_{1,2} = \frac{1}{2} \left[(E_{X_2} + 2E_U)^2 - 2E_U E_{X_2} \mp 
                           \sqrt{(E_{X_2} + 2E_U)^2 (E_{X_2}^2 + 4E_U^2) } \right].
\end{equation}

The magnetization can then be obtained from the free energy:

\begin{equation}
  M = - \frac{\partial F}{\partial B} = M_{nonint}
                   + \frac{\tilde{\lambda} L^2}{\beta} \frac{\partial E_{X_2}}{\partial \phi}
                     \sum_{n} \frac{\frac{\partial f_n}{\partial E_{X_2}} }{1 - \tilde{\lambda}f_n },
\end{equation}
 where $M_{nonint}$ is the contribution from noninteracting
 electrons\cite{Altshuler_Gefen_Imry_91}. We will be interested in the
 second term, which is the fluctuation
 magnetization\cite{Aslamazov_Larkin_68}.

For illustrative purposes, we use the parameters for $Al$ in all our
 numerical calculations, with $\omega_D=34meV$ and
 $\tlambda=0.193$. This leads to a mean-field transition temperature
 $T_{c0}=0.218meV=2.6K$ for an isolated $Al$ quantum dot in the
 absence of magnetic flux. In all our calculations we evaluate
 Matsubara sums with a cutoff $\exp{-|\omega_n|/\omega_D}$. We have
 verified that changing the cutoff does not qualitatively affect our
 results, but only produces small numerical changes.

\begin{figure*}
  \centerline{
    \mbox{\includegraphics[width=3.5in]{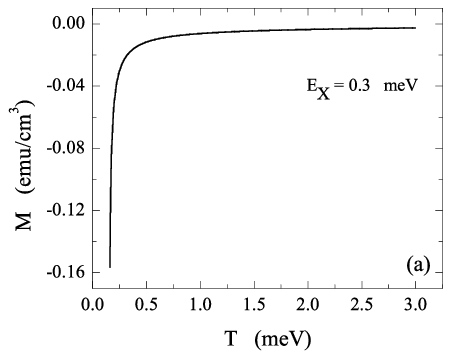}}
    \mbox{\includegraphics[width=3.5in]{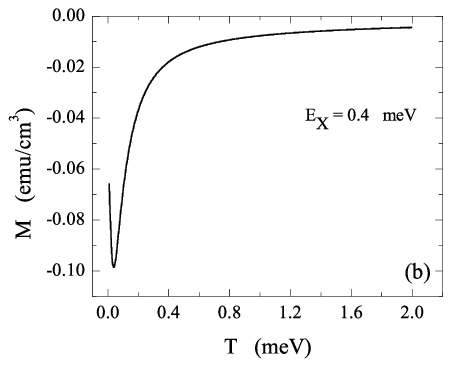}}
  }
  \centerline{
    \mbox{\includegraphics[width=3.5in]{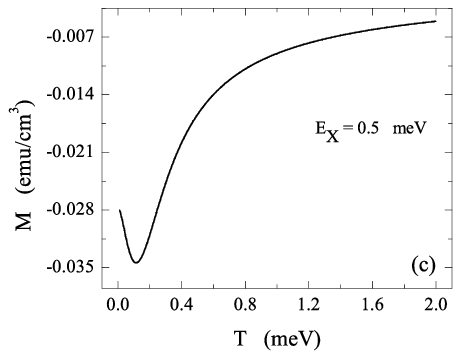}}
    \mbox{\includegraphics[width=3.5in]{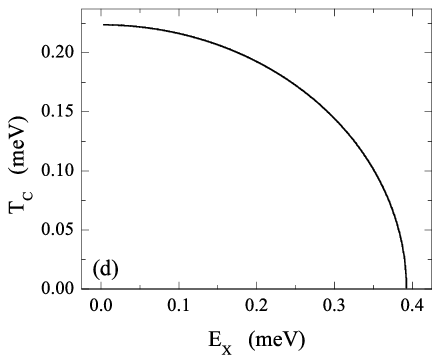}}
  }
  \caption{Magnetization (per unit volume) in a single dot system as a function of temperature for
            different values of crossover parameters $E_{X}$. Panel (d) shows the dependence of the 
            critical temperature on $E_{X}$. }
  \label{View8}
  \end{figure*}

It will be informative to compare the two-dot system with a single dot
subject to an orbital magnetic flux\cite{murthy:article} (see
Fig. \ref{View8}). We draw the reader's attention to two important
features. Firstly, the critical temperature $T_c$ decreases
monotonically with $E_X$, resulting from the fact that time-reversal
breaking disfavors superconductivity. Secondly, the fluctuation
magnetization is always negative, or diamagnetic, resulting from the
fact that the free energy monotonically increases as the orbital flux
increases.

\begin{figure*}
  \centerline{
    \mbox{\includegraphics[width=3.5in]{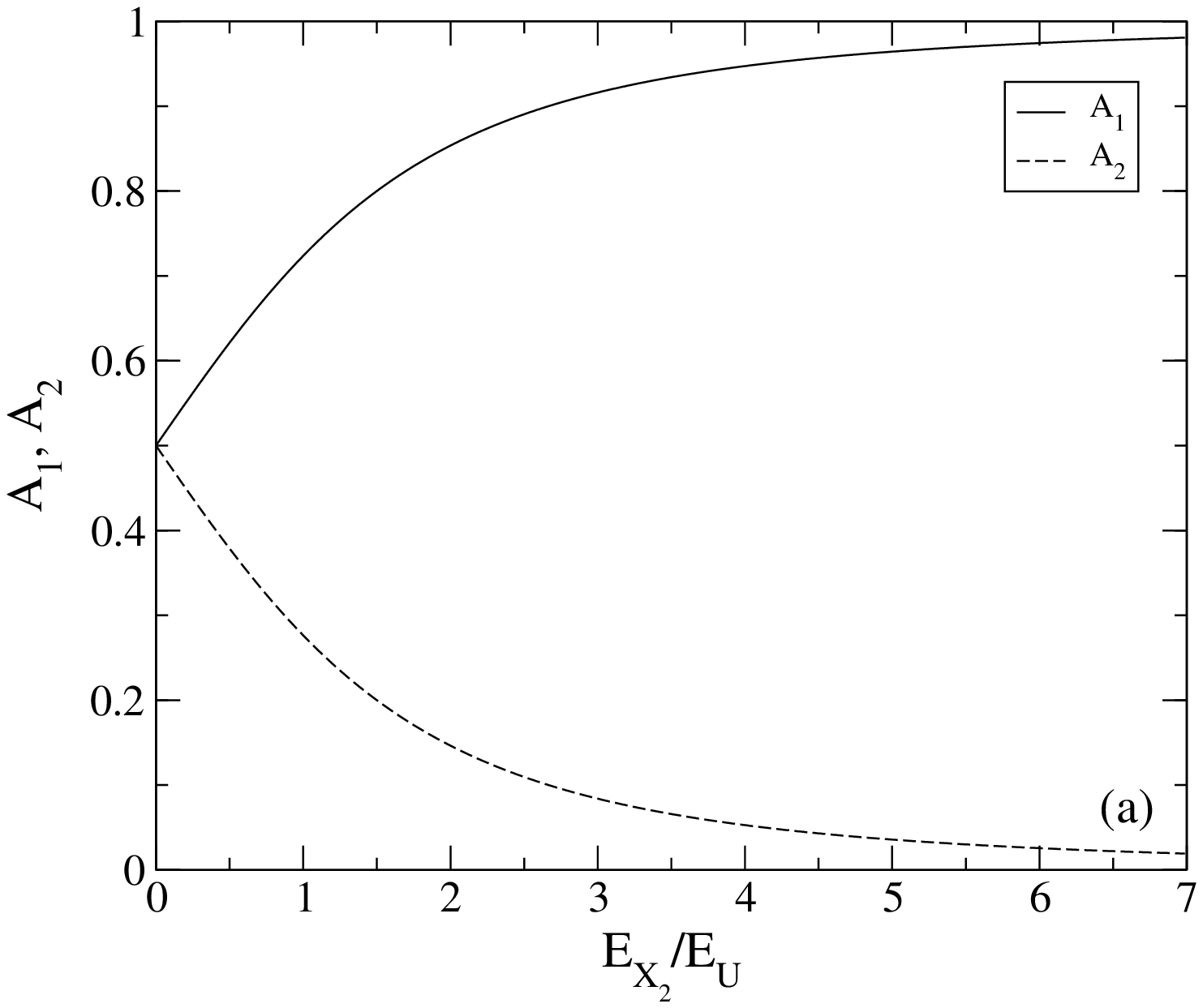}}
    \mbox{\includegraphics[width=3.5in]{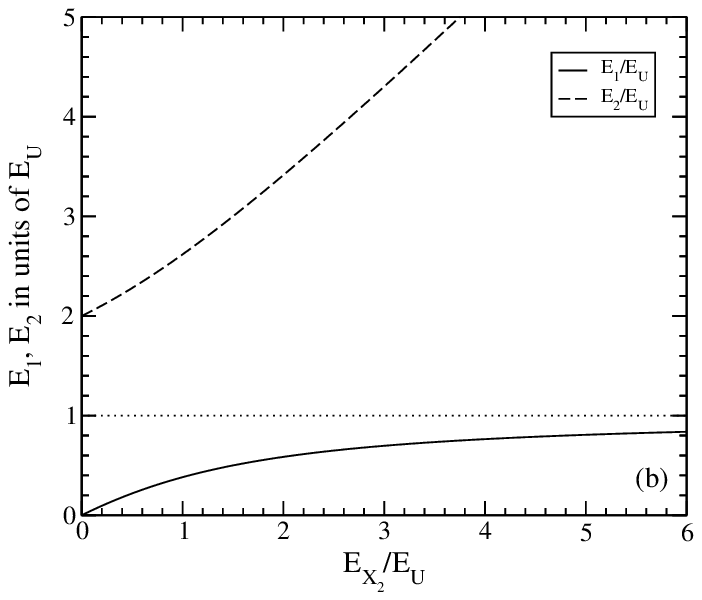}}
  }
  \caption{The behavior of Log coefficients in Eq. \eqref{scalingchi} and 
           $E_1$,$E_2$ as functions of the ratio $E_{X_2}/E_U$.}
  \label{View9}
  \end{figure*}

Now let us turn to our system of two quantum dots coupled by hopping.
Before we carry out a detailed analysis, it is illuminating to inspect
the behavior of $E_{1,2}$ and the coefficients of the two logarithms
in Eq. (\ref{scalingchi}) (which we call $A_{1,2}$) as a function of
$E_{X_2}$. This is shown in Fig. \ref{View9}. $E_1$ tends to $E_{X_2}/2$ for
$E_{X_2}\ll E_U$, and to $E_U$ in the opposite limit $E_{X_2}\gg
E_U$. $E_2$ tends to $E_U$ for $E_{X_2}\ll E_U$, while in the opposite
limit $E_{X_2}\gg E_U$ $E_2\to E_{X_2}$. Both coefficients $A_{1,2}$
start at $\half$ for small $E_{X_2}$. For $E_{X_2}\gg E_U$ $A_1\to 1$,
while $A_2\to0$.

The asymptotic regimes $T,E_{X_2}\ll E_U$ and $T,E_{X_2}\gg E_U$ can
be understood simply. In the first regime, $E_U$ is the largest energy
scale, and far below it the spatial information that there are two
distinct quantum dots is lost. The system behaves like a single large
dot with a smaller ``diluted'' superconducting coupling. On the other
hand, when $T,E_{X_2}\gg E_U$, $A_2$ is vanishingly small, and the
system resembles the isolated first dot with a superconducting
coupling $\tlambda$ but with a crossover energy $E_U$. Note that the
approach of the energies to the asymptotes is slow, so for a
particular value of $E_U$ it may happen that one cannot realistically
approach the asymptotic regime without running into either $\delta$ at
the lower end or $\omega_D$ at the higher end. Finally, one can
envisage situations in which $E_{X_2}\ll E_U$ but $T\ge
E_U$, for which there are no simple pictures.

\begin{figure*}
  \centerline{
    \mbox{\includegraphics[width=3.5in]{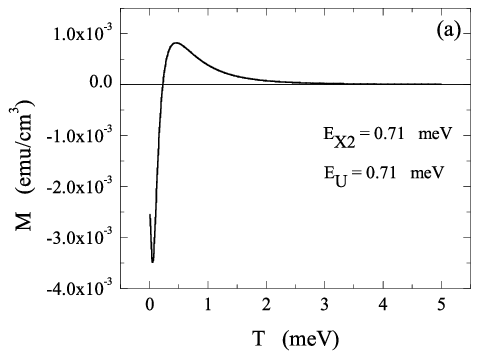}}
    \mbox{\includegraphics[width=3.5in]{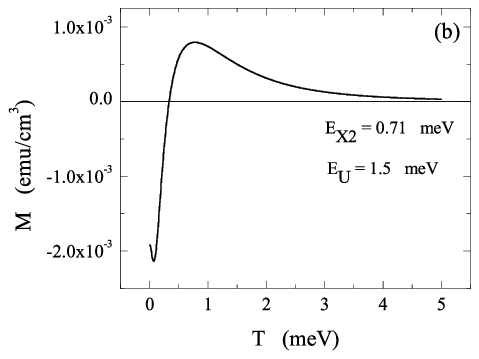}}
  }
  \centerline{
    \mbox{\includegraphics[width=3.5in]{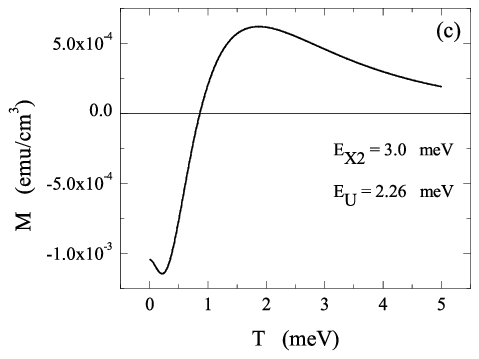}}
    \mbox{\includegraphics[width=3.5in]{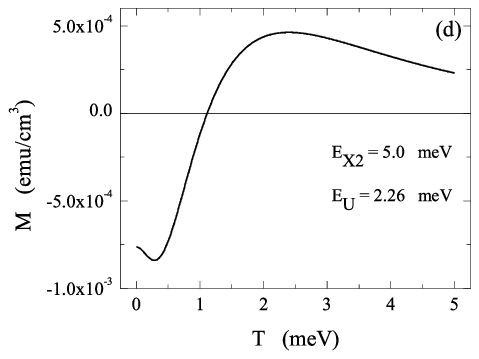}}
  }
  \caption{Magnetization (per unit volume) as a function of
            temperature for different values of crossover parameters
            $E_{X_2}$ and $E_U$. The fluctuation magnetization is
            diamagnetic for low $T$ and paramagnetic for high $T$.}
  \label{View3}
  \end{figure*}

The temperature dependence of magnetization per unit volume for
different values of crossover parameters $E_{X_2}$ and $E_U$
(excluding the part due to noninteracting electrons) is shown in
Fig. \ref{View3}.

\begin{figure*}
  \centerline{
    \mbox{\includegraphics[width=3.5in]{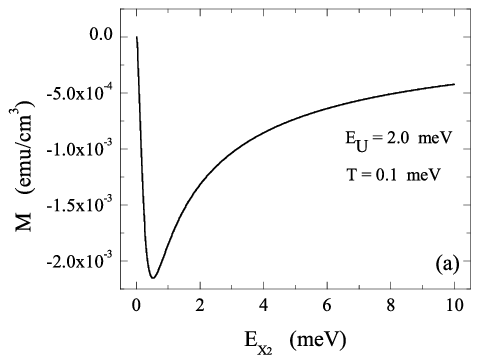}}
    \mbox{\includegraphics[width=3.5in]{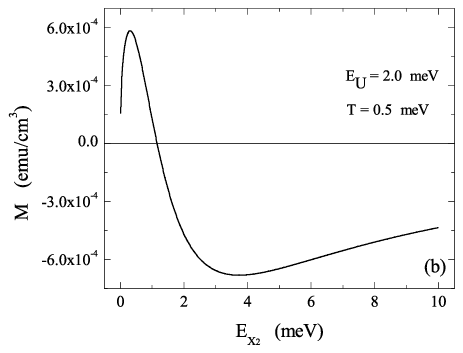}}
  }
  \centerline{
    \mbox{\includegraphics[width=3.5in]{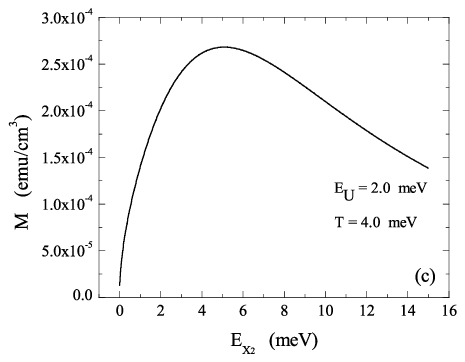}}
  }
  \caption{Fluctuation magnetization in the first dot vs crossover
            parameter $E_{X_2}$ in the second dot for different values
            of temperature. The fluctuation magnetization is
            diamagnetic for low $T$ and paramagnetic for high $T$. }
  \label{View4}
  \end{figure*}

In the range where magnetization changes significantly, the
fluctuation magnetization shows both diamagnetic and paramagnetic
behavior. This is in contrast to the case of a single superconducting
quantum dot subjected to an orbital flux where the fluctuation
magnetization is always diamagnetic (Fig. \ref{View8}). Close to $T=0$ an increase in
temperature makes the fluctuation magnetization more diamagnetic. A
further temperature increase changes the fluctuation magnetization
from diamagnetic to paramagnetic. For large values of temperature the
fluctuation magnetization is paramagnetic and decreasing as $T$
increases.
   Another set of diagrams,  Fig. \ref{View4}, demonstrates the
dependence of the fluctuation magnetization in the first dot on
crossover parameter $E_{X_2}$ in the second dot. Generically, we find
that at low $T$ the fluctuation magnetization is diamagnetic while at
high $T$ it is paramagnetic.

  The variation of crossover energy scales $E_{X_2}$ and $E_U$ does
not change the qualitative behavior of the fluctuation magnetization
as a function of $T$ or $E_{X_2}$. A paramagnetic magnetization is
counterintuitive in superconducting system, because one believes that
``an orbital flux is the enemy of superconductivity'', and therefore
that the free energy must always increase as the orbital flux
increases. This assumption is false for our system. The explanation is
fairly simple, as we will see immediately after the results for $T_c$
have been presented.

\begin{figure*}
  \centerline{ \mbox{\includegraphics[width=3.5in]{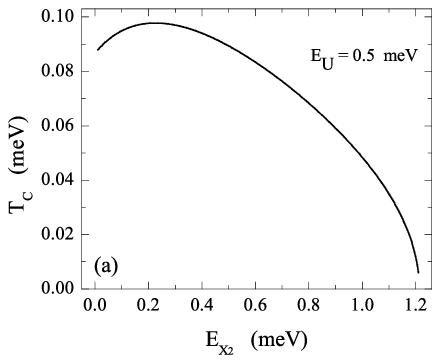}}
    \mbox{\includegraphics[width=3.5in]{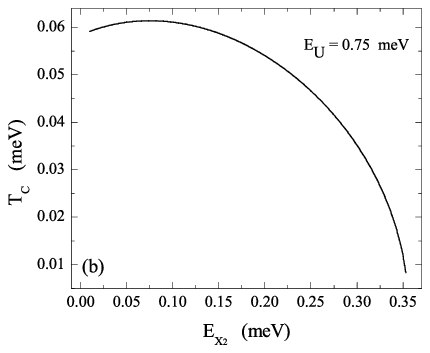}} }
    \centerline{ \mbox{\includegraphics[width=3.5in]{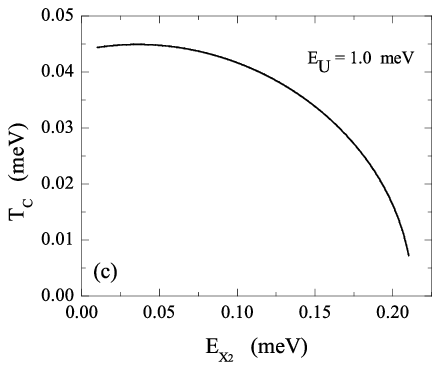}}
    \mbox{\includegraphics[width=3.5in]{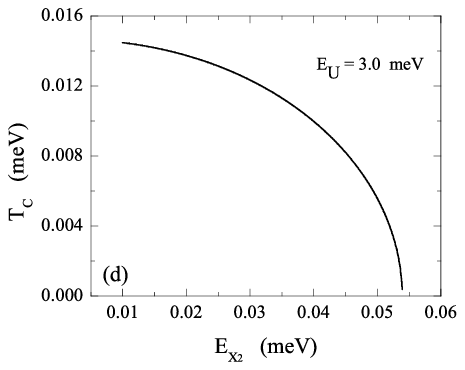}} }
  \caption{Critical temperature as a function of $E_{X_2}$ for several
           intermediate to strong values (compared to $T_{c0}$) of the
           hopping parameter $E_U$. For larger values of $E_{X_2}$
           (not shown on graphs) critical temperature is equal to
           zero.}
  \label{View5}
  \end{figure*}

   The mean-field critical temperature $T_c$ of transition between
normal and superconducting state strongly depends on $E_{X_2}$ and
$E_U$.  As one can see from Fig. \ref{View5}, for very strong hopping
($E_U\gg T_{c0}$) between quantum dots $T_c$ is monotonically
decreasing as $E_{X_2}$ increases. On the other hand, for intermediate
hopping $T_c$ has a maximum as a function of orbital flux, which means
that for small values of orbital magnetic flux $T_c$ {\it increases as
the orbital flux increases}. Finally, when $E_U$ is very weak, $T_c$
{\it monotonically increases} as a function of orbital flux through
the second quantum dot. This is in contrast to the behavior of a
single superconducting quantum dot for which $T_c$ decreases
monotonically as a function of orbital flux.

  \begin{figure*}
  \centerline{
    \mbox{\includegraphics[width=3.5in]{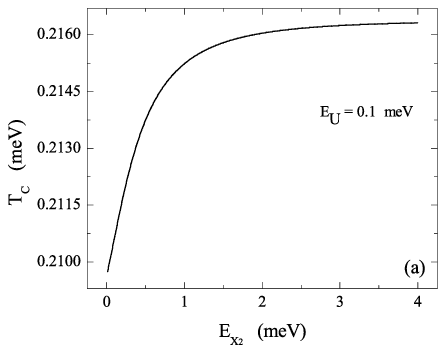}}
    \mbox{\includegraphics[width=3.5in]{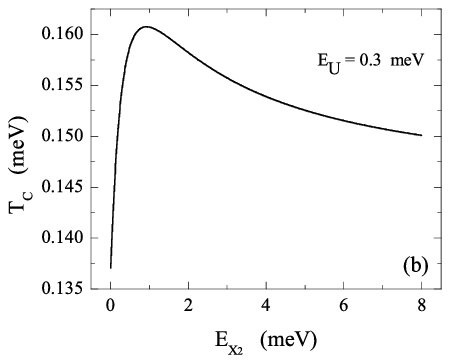}}
  }
  \centerline{
    \mbox{\includegraphics[width=3.5in]{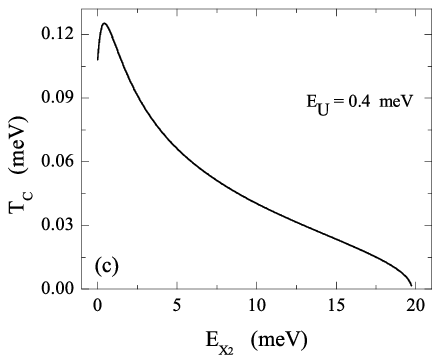}}
    \mbox{\includegraphics[width=3.5in]{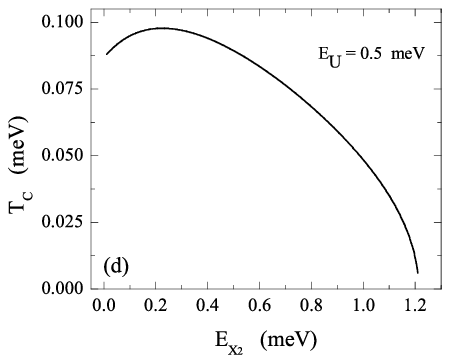}}
  }
  \caption{Behavior of critical temperature $T_C$ as a function of
           $E_{X_2}$ for small to intermediate values (compared to
           $T_{c0}$) of $E_U$.}
  \label{View6}
  \end{figure*}

These counterintuitive phenomena can be understood in terms of the
following cartoon picture. One can think of the two dots as two sites,
each capable of containing a large number of bosons (the fluctuating
pairs). The BCS pairing interaction occurs only on the first
site. When there is no magnetic flux, hopping delocalizes the bosons
between the two sites, leading to a ``dilution'' of the BCS attraction
and a low critical temperature. The effect of the magnetic flux on the
second dot is twofold: (i) Firstly, it gaps the cooperon of the second
dot, which we think of as raising the energy for the bosons to be in
the second dot. (ii) Secondly, by virtue of the interdot hopping, a
small time-reversal symmetry breaking is produced in the first dot,
thereby raising the energy of the bosons there as well. As the flux
through the second dot rises, the bosons prefer to be in the first dot
since they have lower energy there. The more localized the cooper
pairs are in the first dot due to effect (i), the more ``undiluted''
will be the effect of the BCS attraction $\lambda$, and the more
favored will be the superconducting state. However, effect (ii)
produces a time-reversal breaking in the first dot, thus disfavoring
the superconducting state. These two competing effects lead to the
varying behaviors of $T_c$ and the fluctuation magnetization versus
the orbital flux in the second quantum dot. When the hopping between
the quantum dots is weak ($E_U< T_{c0}$), the first effect dominates,
and $T_c$ increases with $E_{X_2}$. When the hopping is stronger
($E_U\simeq T_{c0}$) the first effect dominates at small orbital flux,
and the second at large orbital flux. Finally, at very large hopping
($E_U\gg T_{c0}$), effect (ii) is always dominant.

When considering the magnetization one must take into account the
temperature as well, so the picture is more complex. The general
feature is that effect (i) which tends to localize the pairs in the
first dot also tends to {\it decrease} the interacting free energy of
the system, which leads to a paramagnetic fluctuation
magnetization. Effect (ii), which breaks time-reversal in the first
dot, increases the free energy of the system and thus leads to a
diamagnetic fluctuation magnetization. Based on our results we infer
that at high temperature the coherence of pair hopping is destroyed
leading to more localization in the first quantum dot. The
consequences of high $T$ are thus similar to that of the effect (i): A
lowering of the interacting free energy and a paramagnetic fluctuation
magnetization.

We can make this picture a bit more quantitative for the behavior of
$T_c$ with respect to $E_X$. Consider once more the scaling function of
Eq. (\ref{scalingchi}), which we reproduce here for the reader's
convenience
\begin{multline}
 f_n(E_{X_2},E_U,T) =  \frac{E_U}{2E_1} \frac{E_{X_2}^2 + E_U E_{X_2} - E_1^2}{E_2^2 - E_1^2} 
         \ln \left[ \frac{4(\hbar\omega_D)^2 + \omega_n^2 }{C'/\beta^2 + (E_1+|\omega_n|)^2 } \right]\\
       + \frac{E_U}{2E_2} \frac{E_2^2 - E_{X_2}^2 - E_U E_{X_2} }{E_2^2 - E_1^2}
         \ln \left[ \frac{4(\hbar\omega_D)^2 + \omega_n^2 }{C'/\beta^2 + (E_2+|\omega_n|)^2 } \right].
\label{scalingchi2}\end{multline} 

It is straightforward to show that $f_n$ reaches its maximum value for $\omega_n=0$. The condition for $T_c$ is then 
\begin{equation}
{\tilde{\lambda}}f_0(E_{X_2},E_U,T_c)=1
\end{equation}
Let us first set $E_{X_2}=0$. Let us also call the mean-field critical
temperature of the {\it isolated} first dot in the absence of a
magnetic flux $T_{c0}$ (recall that for the parameters pertinent to
$Al$, $T_{c0}=0.218meV=2.6K$). Now there are two possible limits,
either $E_U\ll T_{c0}$ or $E_U\gg T_{c0}$. In the first case we obtain
\begin{equation}
T_c(E_U)\simeq T_{c0}\bigg(1-{E_U^2\over{\tilde{\lambda}}C' T_{c0}^2}+\cdots\bigg)
\label{tcsmallu}\end{equation}
In the second case, $E_U\gg T_{c0}$, we obtain
\begin{equation}
T_c(E_U)\simeq T_{c0} {\omega_D\over E_U} e^{-1/{\tilde{\lambda}}}
\label{tcbigu}\end{equation}
Note that this can be much smaller than $T_{c0}$ and is an
illustration of the ``dilution'' of the BCS attraction due to the
second dot mentioned earlier. Of course, there will be a smooth
crossover between the expressions of Eq. (\ref{tcsmallu}) and
Eq. (\ref{tcbigu}), so that $T_c$ is always smaller than $T_{c0}$.

Now under the assumption $E_{X_2},\ T_c\ll E_U$ we can solve analytically for $T_c$ to obtain
\begin{equation}
T_c^2(E_{X_2},E_U)\simeq -{E_{X_2}^2\over4C'}+{4\omega_D^4\over C'^2E_U^2} e^{-4/{\tilde{\lambda}}} e^{{2E_{X_2}\over E_U}\big({1\over{\tilde{\lambda}}}-{1\over4}-\ln{\omega_D\over E_U}\big)}
\end{equation}
One can further find the maximum of this expression. It turns out that
$E_U$ has to be larger than a critical value $E_U^*$ for there to be a
maximum.
\begin{equation}
E_U^*=\omega_D e^{({1\over4}-{1\over{\tilde{\lambda}}})}
\end{equation}
For our values of the parameters $\omega_D=34meV$,
${\tilde{\lambda}}=0.193$, we find $E_U^*=0.245meV$. The position of the maximum can now be estimated asymptotically for $E_U>E_U^*$  as 
\begin{equation}
E_{X_2}^*\simeq 16 e^{-1} E_U^* \bigg({E_U^*\over E_U}\bigg)^3 \ln{{E_U\over E_U^*}}
\label{eq:30}
\end{equation}
%

  \begin{figure*}
  \centerline{
    \mbox{\includegraphics[width=5in]{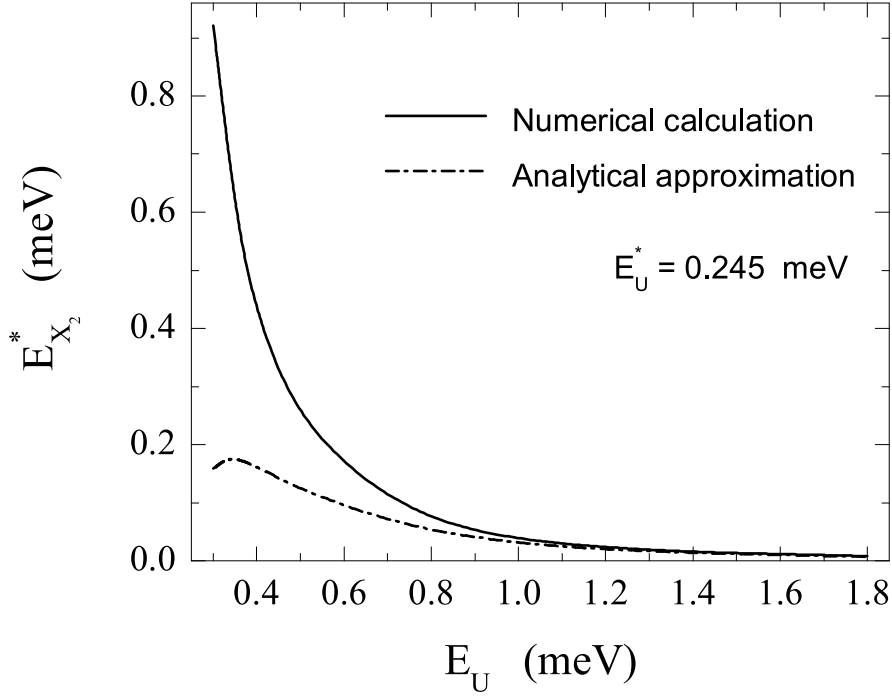}}
  }
  \caption{The behavior of $E^*_{X_2}$ vs $E_U$ for numerical simulation and analytical approximation.}
  \label{View7}
  \end{figure*}

 Fig.\ref{View7} compares the dependence of $E^*_{X_2}$ vs $E_U$ in case of 
 numerical simulation and the one described by Eq. \eqref{eq:30}. For large values of $E_U$ compared to $E_U^*$
the  numerically computed curve  matches the analytical approximation.

\section{Conclusion and Discussion}\label{conclusion}

In writing this paper we began with two objectives. We intended to
compute noninteracting scaling functions in the GOE$\to$GUE crossover
in a system of two dots coupled by hopping, and to use this
information to investigate the properties of an interacting
system\cite{adam:article, Adam_Brouwer_Sharma_03,
Alhassid_Rupp_03:condmat, murthy:article} in the many-body quantum
critical regime\cite{Chakravarty_Halperin_Nelson_89,
Chakravarty_Halperin_Nelson_88, Sachdev:book}.

We have considered a system of two coupled quantum dots, each of which
could have its own time-reversal breaking parameter, coupled by a
bridge which could also have time-reversal breaking. For each
crossover parameter, there is a corresponding crossover energy scale,
which represents the inverse of the time needed for the electron to
``notice'' the presence of that coupling in the Hamiltonian. We have
computed the two-particle Green's functions in the coupled system in a
large-$N$ approximation\cite{aleiner:article}, valid when all energies
of interest are much greater than the mean level spacing. This allows
us to compute the correlations of products of four wavefunctions
belonging to two different energy levels (which have been previously
calculated for a single dot for the pure ensembles by Mirlin using
supersymmetry methods\cite{mirlin_00}, and for the Orthogonal to
Unitary crossover by Adam {\it et al}\cite{adam:article}). The
two-particle Green's function splits naturally into a diffuson part
and a cooperon part. Each of these parts can be represented as
${1\over-i\omega}$ times a scaling function, where $\omega$ represents
the frequency at which the measurement is being performed. For
example, when we use the two-particle Green's function to find the
ensemble average of four wavefunctions belonging to two energies,
$\omega$ is the energy difference between the two states. The
``scaling'' nature of the scaling function is represented by the fact
that it depends only on the ratio of $\omega$ to certain crossover
energy scales. For the diffuson part the crossover energy $E_U$ is
controlled solely by the strength of the hopping between the two dots,
while the scaling function for the cooperon part depends sensitively
on the time-reversal breaking in all three parts of the system.

In the second part of the paper, we consider the case when one of the
dots has an attractive BCS interaction, implying that it would be
superconducting in the mean-field limit at zero temperature if it were
isolated, and the other dot has no electron interactions but is
penetrated by an orbital magnetic flux. The BCS interaction is one
part of the Universal Hamiltonian\cite{Andreev_Kamenev_81,
Brower_Oreg_Halper_99, Baranger_Ulmo_Glazman_00,
Kurland_Aleiner_Altshuler_00}, known to be the correct low-energy
effective theory\cite{Murthy_Mathur_02:paper, Murthy_Shankar_03,
Murthy_Shankar_Herman_Mathur_04} in the renormalization
group\cite{Shankar_94:paper, Shankar_91} sense for weak-coupling and
deep within the Thouless band $|\varepsilon-\varepsilon_F|\ll E_T$. In
order to eliminate complications arising from the charging energy, we
consider a particular geometry with the dots being vertically coupled
and very close together in the vertical direction, as shown in
Fig. \ref{vert-coupled}. Our focus is on the quantum critical
regime\cite{Chakravarty_Halperin_Nelson_89,
Chakravarty_Halperin_Nelson_88, Sachdev:book}, achieved by increasing
either the temperature or the orbital flux through the second dot. The
first dot is coupled by spin-conserving hopping to a second dot on
which the electrons are noninteracting. This coupling always reduces
the critical temperature, due to the ``diluting'' effect of the second
dot, that is, due to the fact that the electrons can now roam over
both dots, while only one of them has a BCS attraction. Thus, the
mean-field critical temperature $T_c$ of the coupled system is always
less than that of the isolated single superconducting dot
$T_{c0}$. This part of the phenomenology is intuitively obvious.

However, when the hopping crossover energy $E_U$ is either weak or of
intermediate strength compared to $T_{c0}$, turning on an orbital flux
in the second dot can lead to a counterintuitive {\it increase} in the
mean-field critical temperature of the entire system. For very weak
hopping, the mean-field $T_c$ monotonically increases with orbital
flux through the second dot, reaching its maximum when the second dot
is fully time-reversal broken. For intermediate hopping strength, the
mean-field $T_c$ initially {\it increases} with increasing orbital
flux to a maximum. Eventually, as the orbital flux, and therefore the
crossover energy corresponding to time-reversal breaking in the second
dot increases, the critical temperature once again decreases. For
strong hopping $E_U\gg T_{c0}$, $T_c$ monotonically decreases as a
function of the orbital flux in the second quantum dot.

We have obtained the detailed dependence of the fluctuation
magnetization in the quantum critical regime as a function of the
dimensionless parameters $T/E_{X_2}$ and $E_{X_2}/E_U$. Once again,
the coupled dot system behaves qualitatively differently from the
single dot in having a {\it paramagnetic} fluctuation magnetization in
broad regimes of $T$, $E_{X_2}$, and $E_U$.

We understand these phenomena qualitatively as the result of two
competing effects of the flux through the second dot. The first effect
is to raise the energy for Cooper pairs in the second dot, thereby
tending to localize the pairs in the first dot, and thus reducing the
``diluting'' effect of the second dot. This first effect tends to
lower the interacting free energy (as a function of orbital flux) and
raise the critical temperature. The second effect is that as the
electrons hop into the second dot and return they carry information
about time-reversal breaking into the first dot, which tends to
increase the free energy (as a function of orbital flux) decrease the
critical temperature. The first effect dominates for weak hopping
and/or high $T$, while the second dominates for strong hopping and/or
low $T$. Intermediate regimes are more complex, and display
nonmonotonic behavior of $T_c$ and the fluctuation magnetization.

It should be emphasized that the quantum critical regime we focus on
is qualitatively different from other {\it single-particle} random
matrix ensembles applicable to a normal mesoscopic system which is
gapless despite being in contact with one or more superconducting
regions\cite{altland-zirnbauer1,altland-zirnbauer2}, either because
the two superconductors have a phase difference of $\pi$ in their
order parameters\cite{altland-zirnbauer1}, or because they are
$d$-wave gapless superconductors\cite{altland-zirnbauer2}. The main
difference is that we investigate and describe an interacting regime,
not a single-particle one. Without the interactions there would be no
fluctuation magnetization.

Let us consider some of the limitations of our work. The biggest
limitation of the noninteracting part of the work is that we have used
the large-$N$ approximation, which means that we cannot trust our
results when the energy scales and/or the frequency of the measurement
becomes comparable to the mean level spacing. When
$\omega\simeq\delta$ the wavefunctions and levels acquire correlations
in the crossover which we have neglected. Another limitation is that
we have used a particular model for the interdot hopping which is
analytically tractable, and is modelled by a Gaussian distribution of
hopping amplitudes. This might be a realistic model in vertically
coupled quantum dots, or where the bridge has a large number of
channels, but will probably fail if the bridge has only a few
channels. These limitations could conceivably be overcome by using
supersymmetric methods\cite{efetov:book,Tschersich_Efetov_00}.

Coming now to the part of our work which deals with interactions, we
have restricted ourselves to the quantum critical regime of the
system, that is, when there is no mean-field BCS gap. Of course, a
finite system cannot undergo spontaneous symmetry-breaking.  However,
in mean-field, one still finds a static BCS gap. The paradox is
resolved by considering phase fluctuations of the order parameter
which restore the broken symmetry\cite{alhassid-fang-schmidt06}. To
systematically investigate this issue one needs to analyze the case
when the bosonic auxiliary field $\s$ in the coupled-dot system
acquires a mean-field expectation value and quantize its phase
fluctuations.

We have also chosen a geometry in which interdot charging effects can
be ignored. However, most experimental systems with superconducting
nanoparticles deal with almost spherical particles.  For two such
nanoparticles coupled by hopping, one cannot ignore charging
effects\cite{Adam_Brouwer_Sharma_03,kamenev-gefen96,efetov-tschersich03,efetov-etal06}.
We expect these to have a nontrivial effect on the mean-field $T_c$
and fluctuation magnetization of the combined system. We defer this
analysis to future work.

There are several other future directions in which this work could be
extended. New symmetry classes\cite{Aleiner_Falko_87, Aleiner_Falko_89_erratum} 
have been discovered recently for two-dimensional
disordered/ballistic-chaotic systems subject to spin-orbit
coupling\cite{Dresselhaus_55, Bychkov_Rashba_84}. In one of these
classes, the spin-orbit coupling is unitarily equivalent to an orbital
flux acting oppositely\cite{Aleiner_Falko_87, Aleiner_Falko_89_erratum} 
on the two eigenstates of a single-particle quantum number algebraically
identical to $\sigma_z$. Due to the unitary transformation, this
quantum number has no simple interpretation in the original
(Orthogonal) basis. However, it is clear that the results of this
paper could be applied, {\it mutatis mutandis}, to two coupled
two-dimensional quantum dots subject to spin-orbit couplings. In
particular, consider the situation where one quantum dot has no
spin-orbit coupling, but does have a Stoner exchange interaction,
while the other dot is noninteracting, but is made of a different
material and has a strong spin-orbit coupling. Work by one of us has
shown\cite{murthy:article} that by tuning the spin-orbit
coupling one can access the quantum critical regime, which is
dominated by many-body quantum fluctuations. The above configuration
offers a way to continuously tune the spin-orbit coupling in the first
dot by changing the strength of the hopping between the dots.

In general, one can imagine a wide range of circumstances where
changing a crossover parameter in one (noninteracting) dot allows one
to softly and tunably break a symmetry in the another (interacting)
dot, thereby allowing one access to a quantum critical regime. We hope
the present work will be useful in exploring such phenomena.


  \begin{acknowledgments}
    The authors would like to thank National Science Foundation for
    partial support under DMR-0311761, and Yoram Alhassid for comments
    on the manuscript. OZ wishes to thank the College of Arts and
    Sciences and the Department of Physics at the University of
    Kentucky for partial support. The authors are grateful to
    O. Korneta for technical help with graphics.
  \end{acknowledgments}
 \appendix
%
%
%

\section{One uncoupled dot}\label{apnx:A}

  In this Appendix we calculate one-particle and two-particle Green's
functions for a single dot undergoing the crossover. The strength of magnetic field inside the
dot is controlled by crossover parameter $X$. The Hamiltonian of the
system in crossover is:

\begin{equation}
  H = \frac{H_S + i X H_A}{\sqrt{1 + X^2}},
\end{equation}
  where $H_{S,A}$ are symmetric and antisymmetric real random matrices with the same variance for matrix 
elements. Normalization $(1+X^2)^{-1/2}$ keeps the mean level spacing $\delta$ fixed as magnetic field 
changes inside the dot.

 We define the retarded one-particle Green's function as follows:
\begin{equation}
   G_{\alpha\beta}^R(E) = \left(\frac{1}{E^+ -H}\right)_{\alpha\beta}
                        = \frac{1}{E^+}\left(I+\frac{H}{E^+}+\frac{H^2}{(E^+)^2}+\ldots
                                       \right)_{\alpha\beta}
                        = \frac{\delta_{\alpha\beta}}{E^+}+\frac{H_{\alpha\beta}}{(E^+)^2}+
                          \frac{H^2_{\alpha\beta}}{(E^+)^3}+\ldots,
\end{equation}
  Here $H$ is a Hamiltonian, and $E^+$ is the energy with infinitely small positive imaginary part 
$E^+ = E + i\eta$.

This series has nice graphical representation:

\begin{equation}
  G^R(E) \  = \ \epsfig{file=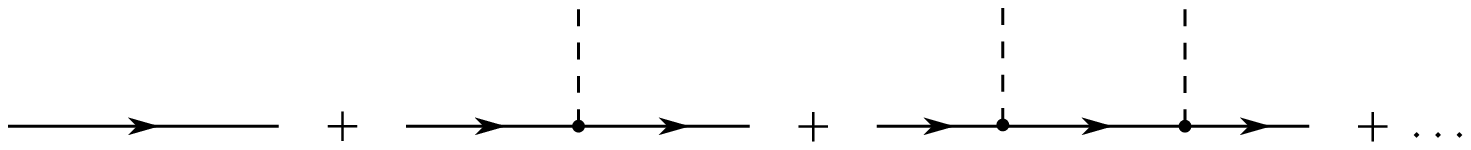, scale=0.7},
\end{equation}
   where straight solid line represents $1/E^+$ and dashed line stands for Hamiltonian.

\begin{equation}
  \frac{1}{E^+} \ =\  \includegraphics{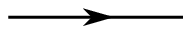},  \hspace{1cm} H \ =\  \includegraphics{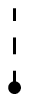}
\end{equation}

Just as in disordered conductor or quantum field theory the target is not the 
Green's function itself, but rather its mean and mean square. We take on random 
matrix ensemble average of $G_{\alpha\beta}$. Such averaging assumes knowledge
of $\langle H^n \rangle$, where angular brackets stand for gaussian ensemble averaging,
and $n = 1, \infty$. For $n=1$ we have $\langle H \rangle = 0$, while for $n=2$
the second moment reads:

\begin{equation}
  \langle H_{\alpha\gamma} H_{\delta\beta}\rangle = 
                                       \frac{\langle H^{s}_{\alpha\gamma}H^{s}_{\delta\beta}\rangle -
                                          X^2 \langle H^{a}_{\alpha\gamma}H^{a}_{\delta\beta}\rangle}{1 + X^2}\\
   = \frac{N\delta^2}{\pi^2} \delta_{\alpha\beta}\delta_{\gamma\delta} + \left( \frac{1-X^2}{1+X^2} \right)
      \frac{N\delta^2}{\pi^2} \delta_{\alpha\delta}\delta_{\gamma\beta}.
\end{equation}

All higher moments of $H$ can be computed using Wick's theorem \cite{stockmann:book}. Thus, the ensemble averaging 
leaves only the terms containing even moments of $H$.
Introducing the notation for $\langle HH \rangle = \includegraphics{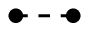}$, we obtain, for the averaged
$G^R$ series:

\begin{equation}
   \langle G^R(E) \rangle = \includegraphics{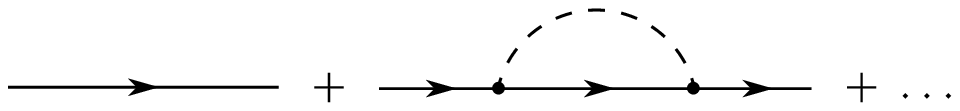}
\label{eq:r1}
\end{equation}

  Then, the expansion \eqref{eq:r1} can be written in a compact form of Dyson equation:

\begin{equation}
        \includegraphics{fig_03_v2.eps}
  \label{eq:31}
\end{equation}

  The bold line denotes the full one-particle Green's function averaged over Gaussian ensemble, and $\Sigma$ is
a self-energy, representing the sum of all topologically different diagrams. The corresponding algebraic expression 
for the Dyson equation can be easily extracted from Eq. \eqref{eq:31} producing:

\begin{equation}
  G_{\alpha\beta} = \sum_{\nu\mu} G_{\alpha\nu} \Sigma_{\nu\mu} \frac{\delta_{\mu\beta}}{E^+}
                  + \frac{\delta_{\alpha\beta}}{E^+},
\label{eq:14}
\end{equation}
   where $G_{\alpha\beta}$ means $\langle G^R_{\alpha\beta}(E) \rangle$. Now, using the fact that 
$G_{\alpha\beta} = G_{\alpha}\delta_{\alpha\beta}$ and 
$\Sigma_{\alpha\beta} = \Sigma_{\alpha}\delta_{\alpha\beta}$ (no summation over $\alpha$ implied), one can solve 
this equation and obtain:

\begin{equation}
  G_{\alpha\beta} = \frac{\delta_{\alpha\beta}}{E^+ - \Sigma}.
\end{equation}

Next we approximate self-energy by the first term in large $N$ approximation:

\begin{equation}
  \Sigma_{\alpha\beta} = \includegraphics{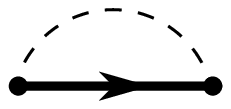} = G\sum_{\gamma} \langle H_{\alpha\gamma} H_{\gamma\beta} \rangle
                       \approx \left( \frac{N\delta}{\pi} \right)^2 \frac{\delta_{\alpha\beta}}{E^+ - \Sigma}.
\label{eq:15}
\end{equation}

Solving Eq. \eqref{eq:15} for the self-energy we determine:

\begin{equation}
  \Sigma = \frac{E}{2} - \frac{i}{2} \sqrt{\left( \frac{2N\delta}{\pi} \right)^2 - E^2}.
\end{equation}

   Consequently, the ensemble average of one-particle Green's function is given by:

\begin{equation}
   \langle G_{\alpha\beta}^R(E) \rangle =
   \frac{\delta_{\alpha\beta}}
        {\frac{E}{2}+\frac{i}{2}\sqrt{\left(\frac{2N\delta}{\pi}\right)^2 - E^2}};
   \hskip 1cm
   \langle G_{\alpha\beta}^A(E) \rangle = \langle G_{\beta\alpha}^R(E) \rangle ^{\ast}.
\end{equation}

  Next, to study the two-particle Green's function we notice that the main contributions come from ladder and maximally 
  crossed diagrams:

\begin{equation}
  \includegraphics{fig_05_v1.eps}
\end{equation}

  Two bold lines on the left side stand for the average two-particle Green's function 
  $\langle G^R(E+\omega) G^A(E) \rangle$. 
The sum of ladder diagrams is described by Bethe-Salpeter equation:

\begin{equation}
        \includegraphics{fig_07_v1.eps}
\end{equation}
  or,
\begin{equation}
   \Pi^{\alpha\beta,D}_{\delta\gamma} = \frac{N\delta^2}{\pi^2}
   \delta_{\alpha\delta}\delta_{\beta\gamma} +
   \left(\frac{N\delta}{\pi}\right)^2
   \frac{\Pi^{\alpha\beta,D}_{\delta\gamma}}{F[E,\omega]},
\end{equation}
   where $\Pi^D$ is a ladder approximation of diffuson part of two-particle Green's function.
   Here $F[E,\omega]$ is a product of two inversed averaged one-particle Green's functions
   and in the limit $\omega \ll N\delta$ is:

\begin{equation}
  F[E,\omega] = \langle G^R(E+\omega) \rangle^{-1} \langle G^A(E) \rangle^{-1} 
                  \approx -\frac{i\omega\delta N}{2\pi} +
                  \left(\frac{N\delta}{\pi}\right)^2.
\end{equation}

  One can solve this equation taking into account 
    $\Pi^{\alpha\beta,D}_{\delta\gamma} = \Pi^D \delta_{\alpha\delta} \delta_{\beta\gamma}$:

\begin{equation}
   \Pi^D = \frac{N\delta^2}{\pi^2}
         \frac{F[E,\omega]}{F[E,\omega] -
         \left(\frac{N\delta}{\pi}\right)^2}.
\end{equation}

   Multiplying $\Pi^D$ by $F^2[E,\omega]$ we arrive at the following expression for the diffuson term:

\begin{equation}
  \langle G^R_{\alpha\gamma}(E+\omega) G^A_{\delta\beta}(E) \rangle_{D} =
  \frac{2\pi}{N^2\delta}
  \frac{\delta_{\alpha\beta}\delta_{\gamma\delta}}{-i\omega}.
\end{equation}

   Then, we turn our attention to the equation for maximally crossed diagrams. We have

\begin{equation}
        \includegraphics{fig_26.eps}
\end{equation}
  and $\Pi^C$ is expressed in terms of $F[E,\omega]$ again:

\begin{equation}
   \Pi^C = \left( \frac{1-X^2}{1+X^2} \right) \frac{N\delta^2}{\pi^2}
         \frac{F[E,\omega]}
              {F[E,\omega] - \frac{1-X^2}{1+X^2}
                             \left(\frac{N\delta}{\pi}\right)^2}.
\end{equation}

   Assuming $X$ to be small compared to unity (weak crossover), we evaluate 
the contribution of maximally crossed diagrams to 
Green's function to get:

\begin{equation}
  \langle G^R_{\alpha\gamma}(E+\omega) G^A_{\delta\beta}(E) \rangle_{C} =
            \frac{2\pi}{N^2\delta}
            \frac{\delta_{\alpha\delta} \delta_{\gamma\beta}}{-i\omega}
                  \frac{1}{1 + i\frac{E_X}{\omega}},
\end{equation}

 where $E_X = 4X^2N\delta/\pi$ is a crossover energy scale. Final expression for the connected part of 
 the two-particle Green's function is:

\begin{equation}
  \langle G^R_{\alpha\gamma}(E + \omega)G^A_{\delta\beta}(E) \rangle = \frac{2\pi}{N^2\delta}
                             \frac{\delta_{\alpha\beta} \delta_{\gamma\delta}}{-i\omega} +
                                                         \frac{2\pi}{N^2\delta}
                             \frac{\delta_{\alpha\delta} \delta_{\gamma\beta}}{-i\omega}
                                 \frac{1}{1 + i\frac{E_X}{\omega}}.
\end{equation}


\section{Two coupled dots}\label{apnx:B}

   This Appendix contains details of the derivation for statistical properties of the Green's functions 
for the two coupled dots connected to each other via hopping bridge $V$. Coupling between dots is 
weak and characterized by dimensionless parameter $U$. For the system of uncoupled dots the Hilbert space 
is a direct sum of spaces for dot 1 and dot 2. Hopping $V$ mixes the states from two spaces. The 
Hamiltonian of the system can be represented as:

\begin{equation}
  H_{tot} = \begin{pmatrix}
               H_{1} & V \\
               V^{\dagger} & H_{2}
            \end{pmatrix}.
\end{equation}

   For $H_{1,2}$ and $V$ we have:

\begin{equation}
  H_n = \frac{ H^S_n + iX_n H^A_n }{\sqrt{1 + X_n^2}}, \ i=1,2; \hspace{1cm}
  V = \frac{ V^R + i\Gamma V^I }{\sqrt{1 + \Gamma^2}}.
\end{equation}

  Here S (A) stands for symmetric (antisymmetric), and R (I) means real (imaginary). Below we use 
Greek indices for dot 1, and Latin indices for dot 2. We also found it convenient to keep 
bandwidth of both dots the same; that is, $N_1\delta_1 = N_2\delta_2$ with $\xi = \delta_1/\delta_2$.

  The following averaged products of matrix elements of $H$ can be obtained:

\begin{equation}
 \begin{split}
  \langle H_{\alpha\gamma} H_{\delta\beta} \rangle &= \frac{N_1\delta_1^2}{\pi^2} 
                 \delta_{\alpha\beta}\delta_{\gamma\delta} + \left( \frac{1-X_1^2}{1+X_1^2} \right)
                  \frac{N_1\delta_1^2}{\pi^2} \delta_{\alpha\delta}\delta_{\gamma\beta} \\
  \langle H_{il} H_{kj} \rangle &= \frac{N_2\delta_2^2}{\pi^2} 
                 \delta_{ij}\delta_{lk} + \left( \frac{1-X_2^2}{1+X_2^2} \right)
                  \frac{N_2\delta_2^2}{\pi^2} \delta_{ik}\delta_{lj},
 \end{split}
\end{equation}

  where $X_1$ and $X_2$ are the crossover parameters in dot 1 and 2. Pairings between $V$ matrix elements are:

\begin{equation}
 \begin{split}
  \langle V_{\alpha i} V_{\beta j}\rangle &= \langle V^{\dagger}_{i\alpha} V^{\dagger}_{j\beta}\rangle =
  \left( \frac{1 - \Gamma^2}{1 + \Gamma^2} \right)
  \frac{\sqrt{N_1N_2}\delta_1\delta_2 U}{\pi^2} \delta_{\alpha\beta} \delta_{ij}\\
  \langle V_{\alpha i} V^{\dagger}_{j\beta}\rangle &= \frac{\sqrt{N_1N_2}\delta_1\delta_2 U}{\pi^2}
                                         \delta_{\alpha\beta}\delta_{ij},
 \end{split}
\end{equation}
  with $\Gamma$ a crossover parameter in hopping bridge.
  Normalization for $V$ pairing is chosen to coincide with that of $\langle HH \rangle$ when 
 $\xi = 1$.

To determine one-particle Green's function we use the system listed in Eq. \eqref{eq:3}.
  The straight and wavy bold lines with arrows represent averaged functions $\langle G^R_1(E) \rangle$, 
  $\langle G^R_2(E) \rangle$ in dot 1 and 2, regular lines represent bare propagators, and the rest of the 
  lines describe pairings between $H_{tot}$ matrix elements. We have:

\begin{equation}
  \begin{split}
    & \includegraphics{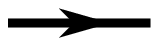} = \langle G^R_1(E) \rangle ; \hspace{1cm} 
      \includegraphics{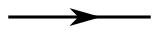} = \frac{1}{E^+} \\
    & \includegraphics{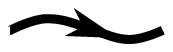} = \langle G^R_2(E) \rangle ; \hspace{1cm} 
      \includegraphics{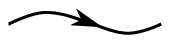} = \frac{1}{E^+}\\
    & \includegraphics{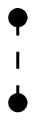} = \langle H_1 H_1 \rangle \hspace{1cm}
      \includegraphics{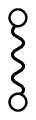} = \langle H_2 H_2 \rangle \hspace{1cm}
      \includegraphics{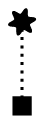} = \langle V V^{\dagger} \rangle.
  \end{split}
\end{equation}

   The corresponding analytical expressions of this system of equations are:
\begin{equation*}
 \begin{split}
   G_1 &= \frac{\Sigma^{11}G_1}{E^+} + \frac{\Sigma^{12}G_1}{E^+} + \frac{1}{E^+}\\
   G_2 &= \frac{\Sigma^{22}G_2}{E^+} + \frac{\Sigma^{21}G_2}{E^+} + \frac{1}{E^+},
 \label{eq:a1}
 \end{split}
\end{equation*}
  with $G_1$ and $G_2$ connected to Green's functions via:
  $\langle G^R_{\alpha\gamma,1}(E) \rangle = G_1\delta_{\alpha\gamma}$, 
  $\langle G^R_{il,2}(E) \rangle = G_2\delta_{il}$ 
  The self-energies $\Sigma^{nm}$ are to be determined using standard procedure \cite{efetov:book}.

 We observe, that the system of two linear equations \eqref{eq:a1} has a solution:

\begin{equation*}
  G_1 = \frac{1}{E^+ - \Sigma^{11} - \Sigma^{12}}, \hspace{1cm}
  G_2 = \frac{1}{E^+ - \Sigma^{22} - \Sigma^{21}}.
\end{equation*}
   Here we approximated self-energies by the first term in
   large N expansion again. In this approximation evaluation of $\Sigma^{nm}$ yields:

\begin{eqnarray*}
    \Sigma_{\alpha\beta}^{11} &=& \Sigma^{11} \delta_{\alpha\beta} = \includegraphics{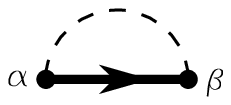} =
        G_1\sum_{\gamma} \langle H_{\alpha\gamma} H_{\gamma\beta} \rangle =
        \left( \frac{N_1\delta_1}{\pi} \right)^2 
        \frac{\delta_{\alpha\beta}}{E^+ - \Sigma^{11} - \Sigma^{12}} \\
    \Sigma_{\alpha\beta}^{12} &=& \Sigma^{12} \delta_{\alpha\beta} = \includegraphics{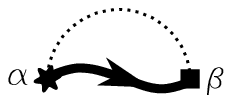} =
        G_2\sum_{i} \langle V_{\alpha i} V^{\dagger}_{i \beta} \rangle =
        \frac{\sqrt{N_1N_2}N_2\delta_1\delta_2 U}{\pi^2}
        \frac{\delta_{\alpha\beta}}{E^+ - \Sigma^{22} - \Sigma^{21}} \\
    \Sigma_{ij}^{22} &=& \includegraphics{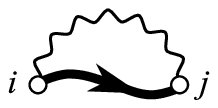} =
        \left( \frac{N_2\delta_2}{\pi} \right)^2
        \frac{\delta_{ij}}{E^+ - \Sigma^{22} - \Sigma^{21}} \\
    \Sigma_{ij}^{21} &=& \includegraphics{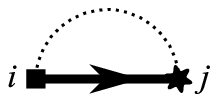} =
        \frac{\sqrt{N_1N_2}N_1\delta_1\delta_2 U}{\pi^2}
        \frac{\delta_{ij}}{E^+ - \Sigma^{11} - \Sigma^{12}}.
\end{eqnarray*}

  Thus, to find all $\Sigma^{nm}$ one needs to solve the following system of equations:

\begin{equation}
\begin{split}
 \Sigma^{11}\left( E^+ - \Sigma^{11} - \Sigma^{12} \right) &= \left( \frac{N_1\delta_1}{\pi}\right)^2 \\
 \Sigma^{12}\left( E^+ - \Sigma^{22} - \Sigma^{21} \right) &= \frac{\sqrt{N_1N_2}N_2\delta_1\delta_2U}{\pi^2} \\
 \Sigma^{22}\left( E^+ - \Sigma^{22} - \Sigma^{21} \right) &= \left( \frac{N_2\delta_2}{\pi}\right)^2 \\
 \Sigma^{21}\left( E^+ - \Sigma^{11} - \Sigma^{12} \right) &= \frac{\sqrt{N_1N_2}N_1\delta_1\delta_2U}{\pi^2}.
\end{split}
\label{eq:4}
\end{equation}

   Observing that $\Sigma^{21} = U\Sigma^{11}/\sqrt{\xi}$ and $\Sigma^{12} = U\sqrt{\xi}\Sigma^{11}$
we decouple the system given in Eq. \eqref{eq:4}. For example, the pair of first and third equations 
can be rewritten as:

\begin{equation}
 \begin{split}
   (\Sigma^{11})^2 - E\Sigma^{11} + U\sqrt{\xi}\Sigma^{11}\Sigma^{22} &=
                                       -\left(\frac{N_1\delta_1}{\pi}\right)^2 \\
   (\Sigma^{22})^2 - E\Sigma^{22} +  \frac{U}{\sqrt{\xi}}\Sigma^{11}\Sigma^{22} &=
                                       -\left(\frac{N_2\delta_2}{\pi}\right)^2.
 \end{split}
\label{eq:5}
\end{equation}

  For weak coupling the solution can be found by expanding self-energies 
$\Sigma^{11}$ and $\Sigma^{22}$ 
in series in $U$. Taking the solution for single dot as zero approximation 
(below all the solutions for 
the uncoupled dot will be marked with subscript $0$) we get

\begin{eqnarray}
   \Sigma^{11} &=& \Sigma_0^{11} + U\Sigma_1^{11}\\
   \Sigma^{22} &=& \Sigma_0^{22} + U\Sigma_1^{22}.
\label{eq:a2}
\end{eqnarray}

  Note that $N_1\delta_1 = N_2\delta_2$, and $\Sigma^{11}_0 = \Sigma^{22}_0 \equiv \Sigma_0$.

  Plugging into the right hand side of Eq. \eqref{eq:a2} in system \eqref{eq:5} we arrive at:

\begin{equation}
 \begin{split}
  \Sigma^{11} &= \Sigma_0 \left( 1 + U\sqrt{\xi} \frac{\Sigma_0}{E^+ - 2 \Sigma_0}\right) \\
  \Sigma^{22} &= \Sigma_0 \left( 1 + \frac{U}{\sqrt{\xi}} \frac{\Sigma_0}{E^+ - 2 \Sigma_0}\right) \\
  \Sigma^{21} &= \frac{U}{\sqrt{\xi}} \Sigma^{11} \\
  \Sigma^{12} &=  U\sqrt{\xi} \Sigma^{22}.
 \end{split}
\end{equation}

   Neglecting the higher powers in $U$ for one-particle Green's functions we finally arrive at
   the following expressions for the single particle Green's functions:

\begin{eqnarray*}
  \langle G^R_{\alpha\beta,1}(E)\rangle &=& \frac{\langle G^R_{\alpha\beta,0}(E)\rangle}{1-U\sqrt{\xi}\frac{\Sigma_0}{E-2\Sigma_0} }
                           = \frac{\delta_{\alpha\beta}}
                                          {\left( \frac{N_1\delta_1}{\pi} \right)
                                           \left[ \epsilon + i\sqrt{1-\epsilon^2} \right]}
                             \frac{1}{ \left[ 1+\frac{U\sqrt{\xi}}{2}
                             \left( 1+i\frac{\epsilon}{\sqrt{1-\epsilon^2}} \right) \right] } \\
  \langle G^R_{ij,2}(E)\rangle &=& \frac{\langle G^R_{ij,0}(E)\rangle}{1-\frac{U}{\sqrt{\xi}}\frac{\Sigma_0}{E-2\Sigma_0} }
                           = \frac{\delta_{ij}}
                                  {\left( \frac{N_2\delta_2}{\pi} \right)
                                   \left[ \epsilon + i\sqrt{1-\epsilon^2} \right]}
                             \frac{1}{\left[ 1+\frac{U}{2\sqrt{\xi}}
                             \left( 1+i\frac{\epsilon}{\sqrt{1-\epsilon^2}} \right) \right] },
\end{eqnarray*}
  where $\epsilon = \pi E/2N\delta$.

   Now we switch our attention to the calculational procedure for the average of the two-particle 
   Green's functions
 $\langle G^R_{\alpha\gamma,1}(E+\omega) G^A_{\delta\beta,1}(E) \rangle$  and
 $\langle G^R_{il,2}(E+\omega) G^A_{kj,2}(E) \rangle$.
 In the limit of large $N_1$ and $N_2$ ladder and maximally crossed diagrams contribute the most.
For ladder diagrams we obtain the system of Bethe-Salpeter equations (see Eq. \eqref{eq:6}).
 Here we used the following notation:
\begin{equation}
  \begin{split}
   \includegraphics{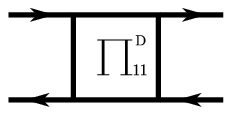} &= \langle G^R_{11}(E+\omega) G^A_{11}(E) \rangle, \hspace{1cm}
   \includegraphics{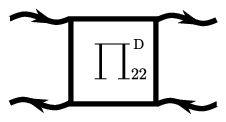}  = \langle G^R_{22}(E+\omega) G^A_{22}(E) \rangle \\
   \includegraphics{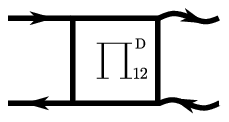} &= \langle G^R_{12}(E+\omega) G^A_{21}(E) \rangle, \hspace{1cm}
   \includegraphics{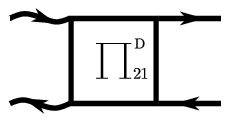}  = \langle G^R_{21}(E+\omega) G^A_{12}(E) \rangle
  \end{split}
\end{equation}

   For the diffuson $\Pi^D_{nm}$ the system of algebraic equations reeds:

\begin{equation}
 \begin{split}
   \Pi^D_{11} &= \frac{N_1\delta_1^2}{\pi^2} + \frac{N_1\delta_1^2}{\pi^2}
              \frac{N_1 \Pi^D_{11}}{F_1[E,\omega]}+
              \frac{\sqrt{N_1N_2}\delta_1\delta_2 U}{\pi^2}
              \frac{N_2 \Pi^D_{21}}{F_2[E,\omega]}\\
   \Pi^D_{22} &= \frac{N_2\delta_2^2}{\pi^2} + \frac{N_2\delta_2^2}{\pi^2}
              \frac{N_2 \Pi^D_{22}}{F_2[E,\omega]}+
              \frac{\sqrt{N_1N_2}\delta_1\delta_2 U}{\pi^2}
              \frac{N_1 \Pi^D_{12}}{F_1[E,\omega]}\\
   \Pi^D_{12} &= \frac{\sqrt{N_1N_2}\delta_1\delta_2 U}{\pi^2} +
              \frac{N_1\delta_1^2}{\pi^2}
              \frac{N_1 \Pi^D_{12}}{F_1[E,\omega]}+
              \frac{\sqrt{N_1N_2}\delta_1\delta_2 U}{\pi^2}
              \frac{N_2 \Pi^D_{22}}{F_2[E,\omega]}\\
   \Pi^D_{21} &= \frac{\sqrt{N_1N_2}\delta_1\delta_2 U}{\pi^2} +
              \frac{N_2\delta_2^2}{\pi^2}
              \frac{N_2 \Pi^D_{21}}{F_2[E,\omega]}+
              \frac{\sqrt{N_1N_2}\delta_1\delta_2 U}{\pi^2}
              \frac{N_1 \Pi^D_{11}}{F_1[E,\omega]},
 \end{split}
\label{eq:7}
\end{equation}
  where $F_1[E,\omega]$ and $F_2[E,\omega]$ are defined as products of inverse averaged one-particle
Green's functions in the first and second dots respectively.
For small values of $U$ and $\omega$ these functions can be approximated as
follows:

\begin{equation}
 \begin{split}
  F_1[E,\omega] &= \langle G^R_1(E+\omega)\rangle^{-1} \langle G^A_1(E)\rangle^{-1} 
         \approx \left(\frac{N_1\delta_1}{\pi}\right)^2
         \left[1+\sqrt{\xi}U-i\tilde{\omega}\right] \\
  F_2[E,\omega] &= \langle G^R_2(E+\omega)\rangle^{-1} \langle G^A_2(E)\rangle^{-1}
                \approx \left(\frac{N_2\delta_2}{\pi}\right)^2
         \left[1+\frac{U}{\sqrt{\xi}}-i\tilde{\omega}\right],
 \end{split}
\end{equation}
  where $\tilde{\omega} = \pi\omega/2N\delta$. The system of four equations given by 
the Eq. \eqref{eq:7} can be decoupled into the two systems of two equations each. 
To determine $\Pi^D_{11}$ one
solves the system of the first and the last equations of Eq. \eqref{eq:7} to get:

\begin{equation}
 \begin{split}
   \left(1-\frac{\left(\frac{N_1\delta_1}{\pi}\right)^2}
                {F_1(E,\omega)}\right)\Pi^D_{11} -
   \frac{\sqrt{N_1N_2}N_2\delta_1\delta_2 U}{\pi^2}
   \frac{\Pi^D_{21}}{F_2(E,\omega)} &= \frac{N_1\delta_1^2}{\pi^2}\\
   \left(1-\frac{\left(\frac{N_2\delta_2}{\pi}\right)^2}
                {F_2(E,\omega)}\right)\Pi^D_{21} -
   \frac{\sqrt{N_1N_2}N_1\delta_1\delta_2 U}{\pi^2}
   \frac{\Pi^D_{11}}{F_1(E,\omega)} &=
   \frac{\sqrt{N_1N_2}\delta_1\delta_2 U}{\pi^2}.
 \end{split}
\label{eq:8}
\end{equation}

   Then, solving the resulting system (Eq. \eqref{eq:8}) and attaching external lines one obtains expression for the two-particle 
Green's function in dot 1:

\begin{equation}
  \langle G^R_{\alpha\gamma,1}(E+\omega) G^A_{\delta\beta,1}(E) \rangle_{D} =
                   \frac{2\pi}{N_1^2\delta_1} \frac{\delta_{\alpha\beta}\delta_{\gamma\delta}}{-i\omega}
                   \frac{1+\frac{i}{\sqrt{\xi}} \frac{E_U}{\omega}}
                        {1+i(\sqrt{\xi} + \frac{1}{\sqrt{\xi}}) \frac{E_U}{\omega}}.
\end{equation}

  The corresponding correlator for dot 2 is readily obtained as well:

\begin{equation}
      \langle G^R_{il,2}(E+\omega) G^A_{kj,2}(E) \rangle_{D} =
                   \frac{2\pi}{N_2^2\delta_2} \frac{\delta_{ij}\delta_{lk}}{-i\omega}
                   \frac{1+i\sqrt{\xi} \frac{E_U}{\omega}}
                        {1+i(\sqrt{\xi} + \frac{1}{\sqrt{\xi}}) \frac{E_U}{\omega}}.
\end{equation}

For the second part of the Green's function (which is the sum of maximally crossed diagrams) the
system of equations is described by Eq. \eqref{eq:9}.
Transforming this graphical system into the algebraic one, we get:

\begin{equation}
 \begin{split}
   \Pi^C_{11} &= \left( \frac{1-X^2_1}{1+X^2_1} \right) \frac{N_1\delta_1^2}{\pi^2} +
                \left( \frac{1-X^2_1}{1+X^2_1} \right) \frac{N_1\delta_1^2}{\pi^2}
                \frac{N_1 \Pi^C_{11}}{F_1[E,\omega]}+
                \left( \frac{1-\Gamma^2}{1+\Gamma^2} \right)
                \frac{\sqrt{N_1N_2}\delta_1\delta_2 U}{\pi^2}
                \frac{N_2 \Pi^C_{21}}{F_2[E,\omega]}\\
   \Pi^C_{22} &= \left( \frac{1-X^2_2}{1+X^2_2} \right) \frac{N_2\delta_2^2}{\pi^2} +
                \left( \frac{1-X^2_2}{1+X^2_2} \right) \frac{N_2\delta_2^2}{\pi^2}
                \frac{N_2 \Pi^C_{22}}{F_2[E,\omega]}+
                \left( \frac{1-\Gamma^2}{1+\Gamma^2} \right)
                \frac{\sqrt{N_1N_2}\delta_1\delta_2 U}{\pi^2}
              \frac{N_1 \Pi^C_{12}}{F_1[E,\omega]}\\
   \Pi^C_{12} &= \left( \frac{1-\Gamma^2}{1+\Gamma^2} \right)
                \frac{\sqrt{N_1N_2}\delta_1\delta_2 U}{\pi^2} +
                \left( \frac{1-\Gamma^2}{1+\Gamma^2} \right)
                \frac{\sqrt{N_1N_2}\delta_1\delta_2 U}{\pi^2}
                \frac{N_2 \Pi^C_{22}}{F_2[E,\omega]} +
                \left( \frac{1-X^2_1}{1+X^2_1} \right)
                \frac{N_1\delta_1^2}{\pi^2}
                \frac{N_1 \Pi^C_{12}}{F_1[E,\omega]}\\
   \Pi^C_{21} &= \left( \frac{1-\Gamma^2}{1+\Gamma^2} \right)
                \frac{\sqrt{N_1N_2}\delta_1\delta_2 U}{\pi^2} +
                \left( \frac{1-\Gamma^2}{1+\Gamma^2} \right)
                \frac{\sqrt{N_1N_2}\delta_1\delta_2 U}{\pi^2}
                \frac{N_1 \Pi^C_{11}}{F_1[E,\omega]} +
                \left( \frac{1-X^2_2}{1+X^2_2} \right)
                \frac{N_2\delta_2^2}{\pi^2}
                \frac{N_2 \Pi^C_{21}}{F_2[E,\omega]}.
 \end{split}
\end{equation}

Once again, the system at hand breaks into systems of two equations each. We proceed by combining 
the first and the last equations to obtain:

\begin{equation}
 \begin{split}
      \left[ 1 - \left( \frac{1-X^2_1}{1+X^2_1} \right)
      \frac{\left( \frac{N_1\delta_1}{\pi} \right)^2}
           {F_1[E,\omega]} \right] \Pi^C_{11} -
      \left( \frac{1-\Gamma^2}{1+\Gamma^2} \right)
      \frac{\sqrt{N_1N_2}N_2\delta_1\delta_2 U}{\pi^2}
      \frac{\Pi^C_{21}}{F_2[E,\omega]} &=
      \left( \frac{1-X^2_1}{1+X^2_1} \right)
      \frac{N_1\delta_1^2}{\pi^2}\\
    - \left( \frac{1-\Gamma^2}{1+\Gamma^2} \right)
      \frac{\sqrt{N_1N_2}N_1\delta_1\delta_2 U}{\pi^2}
      \frac{\Pi^C_{11}}{F_1[E,\omega]} +
      \left[ 1 - \left( \frac{1-X^2_2}{1+X^2_2} \right)
      \frac{\left( \frac{N_2\delta_2}{\pi} \right)^2}
           {F_2[E,\omega]} \right] \Pi^C_{21} &=
      \left( \frac{1-\Gamma^2}{1+\Gamma^2} \right)
      \frac{\sqrt{N_1N_2}\delta_1\delta_2 U}{\pi^2}.
 \end{split}
\end{equation}

Now we can construct approximations for the expressions, containing crossover parameters. 
For example, for small values of $X$ and $\Gamma$ the solution for $\Pi^{C}_{11}$ is expressed as follows:

\begin{equation}
  \Pi^C_{11} = \frac{N_1\delta_1^2}{\pi^2}
             \frac{(1-2X_1^2) (\frac{U}{\sqrt{\xi}}-i\tilde{\omega}+2X_2^2) +
                   (1-4\Gamma^2) U^2}
                  {(\sqrt{\xi}U-i\tilde{\omega}+2X_1^2) (\frac{U}{\sqrt{\xi}} -
                    i\tilde{\omega}+2X_2^2) - (1-4\Gamma^2) U^2}.
\end{equation}

  Next, introducing crossover energy scales:

\begin{equation}
  E_{X} = 4 X^2 \frac{N\delta}{\pi} \hspace{1cm} E_U = 2U\frac{N \delta}{\pi} \hspace{1cm}
  E_{\Gamma} = \frac{4\Gamma^2 E_U}{\sqrt{\xi} + \frac{1}{\sqrt{\xi}}}
\end{equation}
  we obtain the solution for $\Pi^C_{11}$ in the following form:

\begin{equation}
  \Pi^C_{11} = \frac{N_1\delta_1^2}{\pi^2} \frac{1}{-i\tilde{\omega}}
             \frac{1-\frac{E_U}{\sqrt{\xi} i\omega}-\frac{E_{X_2}}{i\omega} }
                  {1 - \frac{E_{X_1}+E_{X_2}}{i\omega} + \frac{E_{X_1} E_{X_2}}{(i\omega)^2} +
                   \frac{E_{X_1} E_U}{\sqrt{\xi} (i\omega)^2} +
                   \frac{\sqrt{\xi} E_{X_2} E_U}{(i\omega)^2} +
                   \left( \sqrt{\xi} + \frac{1}{\sqrt{\xi}} \right) \frac{E_U}{i\omega}
                   \left( \frac{E_{\Gamma}}{i\omega} - 1 \right) }.
\end{equation}

   Then, adding external lines to $\Pi^C_{11}$ for Green's function we get:

\begin{equation}
  \begin{split}
  \langle & G^R_{\alpha\gamma,1}(E+\omega) G^A_{\delta\beta,1}(E) \rangle_{C} = \\
                &  \frac{2\pi}{N_1^2\delta_1} \frac{\delta_{\alpha\delta}\delta_{\gamma\beta}}{-i\omega}
                   \frac{1+\frac{i}{\sqrt{\xi}} \frac{E_U}{\omega} + i\frac{E_{X_2}}{\omega} }
                        {1+i\frac{E_{X_1}+E_{X_2}}{\omega} - \frac{E_{X_1} E_{X_2}}{\omega^2} -
                         \frac{E_{X_1}E_U}{\sqrt{\xi}\omega^2} - \frac{\sqrt{\xi}E_{X_2}E_U}{\omega^2} +
                         i\left( \sqrt{\xi} + \frac{1}{\sqrt{\xi}} \right) \frac{E_U}{\omega}
                         \left( 1+i\frac{E_{\Gamma}}{\omega} \right) }.
  \end{split}
\end{equation}

 Similar manipulations for the corresponding correlator of Green's functions for the 
second room result in:

\begin{equation}
  \begin{split}
  \langle & G^R_{il,2}(E+\omega) G^A_{kj,2}(E) \rangle_{C} = \\
                &  \frac{2\pi}{N_2^2\delta_2} \frac{\delta_{ik}\delta_{lj}}{-i\omega}
                   \frac{1 + i\sqrt{\xi} \frac{E_U}{\omega} + i\frac{E_{X_1}}{\omega} }
                        {1+i\frac{E_{X_1}+E_{X_2}}{\omega} - \frac{E_{X_1} E_{X_2}}{\omega^2} -
                         \frac{E_{X_1}E_U}{\sqrt{\xi}\omega^2} - \frac{\sqrt{\xi}E_{X_2}E_U}{\omega^2} +
                         i\left( \sqrt{\xi} + \frac{1}{\sqrt{\xi}} \right) \frac{E_U}{\omega}
                         \left( 1+i\frac{E_{\Gamma}}{\omega} \right) }.
  \end{split}
\end{equation}

  Finally, the connected part of the total two-particle Green's function is obtained as a sum of diffuson and 
cooperon parts, yielding:

\begin{equation}
  \begin{split}
  \langle & G^R_{\alpha\gamma,1}(E+\omega) G^A_{\delta\beta,1}(E) \rangle =
             \frac{2\pi}{N_1^2\delta_1} \frac{\delta_{\alpha\beta}\delta_{\gamma\delta}}{-i\omega}
             \frac{1+\frac{i}{\sqrt{\xi}} \frac{E_U}{\omega}}
                  {1+i(\sqrt{\xi} + \frac{1}{\sqrt{\xi}}) \frac{E_U}{\omega}} + \\
                &  \frac{2\pi}{N_1^2\delta_1} \frac{\delta_{\alpha\delta}\delta_{\gamma\beta}}{-i\omega}
                   \frac{1+\frac{i}{\sqrt{\xi}} \frac{E_U}{\omega} + i\frac{E_{X_2}}{\omega} }
                        {1+i\frac{E_{X_1}+E_{X_2}}{\omega} - \frac{E_{X_1} E_{X_2}}{\omega^2} -
                         \frac{E_{X_1}E_U}{\sqrt{\xi}\omega^2} - \frac{\sqrt{\xi}E_{X_2}E_U}{\omega^2} +
                         i\left( \sqrt{\xi} + \frac{1}{\sqrt{\xi}} \right) \frac{E_U}{\omega}
                         \left( 1+i\frac{E_{\Gamma}}{\omega} \right) }
  \end{split}
\end{equation}
\begin{equation}
  \begin{split}
  \langle & G^R_{il,2}(E+\omega) G^A_{kj,2}(E) \rangle =
          \frac{2\pi}{N_2^2\delta_2} \frac{\delta_{ij}\delta_{lk}}{-i\omega}
          \frac{1+i\sqrt{\xi} \frac{E_U}{\omega}}
               {1+i(\sqrt{\xi} + \frac{1}{\sqrt{\xi}}) \frac{E_U}{\omega}} + \\
                &  \frac{2\pi}{N_2^2\delta_2} \frac{\delta_{ik}\delta_{lj}}{-i\omega}
                   \frac{1 + i\sqrt{\xi} \frac{E_U}{\omega} + i\frac{E_{X_1}}{\omega} }
                        {1+i\frac{E_{X_1}+E_{X_2}}{\omega} - \frac{E_{X_1} E_{X_2}}{\omega^2} -
                         \frac{E_{X_1}E_U}{\sqrt{\xi}\omega^2} - \frac{\sqrt{\xi}E_{X_2}E_U}{\omega^2} +
                         i\left( \sqrt{\xi} + \frac{1}{\sqrt{\xi}} \right) \frac{E_U}{\omega}
                         \left( 1+i\frac{E_{\Gamma}}{\omega} \right) }.
  \end{split}
\end{equation}
%
%


\section{Fourier Transform of two-particle Green's function}\label{apnx:C}

To be able to study temporal behavior of electrons in the rmt system we introduce the 
Fourier transform of two-particle Green's function. We define it via the following integral:

\begin{equation}
  \langle G^R_{\alpha\gamma}(t) G^A_{\delta\beta}(t)\rangle =
            \frac{1}{(2\pi)^2} \int_{-\infty}^{+\infty} \exp^{-i\omega t}
            \langle G^R_{\alpha\gamma}(E+\omega) G^A_{\delta\beta}(E)\rangle d\omega dE.
\end{equation}

To get the correct behavior of the diffuson part for small $\omega$, we replace $1/\omega$ by
$\omega/(\omega^2 + \eta^2)$, where $\eta$ is infinitesimal positive number. Now we introduce 
for dot 1:

\begin{multline}
  f_D(\omega) = \frac{2\pi}{N_1^2\delta_1}\frac{\delta_{\alpha\beta} \delta_{\gamma\delta}}{-i\omega}
                \frac{1+\frac{i}{\sqrt{\xi}} \frac{E_U}{\omega}}
                     { 1 + i\left( \sqrt{\xi} + \frac{1}{\sqrt{\xi}} \right) \frac{E_U}{\omega}}
             \rightarrow
                \delta_{\alpha\beta}\delta_{\gamma\delta}
                \frac{2\pi}{N_1^2\delta_1} \frac{i\omega}{\omega^2 + \eta^2}
                \frac{\omega + \frac{i}{\sqrt{\xi}}E_U }
                     {\omega + i\left( \sqrt{\xi} + \frac{1}{\sqrt{\xi}} \right)E_U } \\
           = \delta_{\alpha\beta}\delta_{\gamma\delta}
              \frac{2\pi}{N_1^2\delta_1} \frac{i\omega \left( \omega + \frac{i}{\sqrt{\xi}} E_U \right)}
                  {(\omega - i\eta)(\omega + i\eta)(\omega+i(\sqrt{\xi} +
                      \frac{1}{\sqrt{\xi}})E_U )}.
\end{multline}

The Fourier transform of this diffuson term gives:

\begin{equation}
  f_D(t) = \frac{1}{2\pi}\int_{-\infty}^{+\infty} \exp(-i\omega t) f_D(\omega) d\omega.
\end{equation}

Next steps are the standard steps of integration in complex plane.
For $t > 0$ one closes contour in lowerhalf plane. One root is located in upper half plane and
two more are located in lower half plane. The integration yields:

\begin{equation}
  f_D(t) = \delta_{\alpha\beta}\delta_{\gamma\delta} \frac{2\pi}{N_1^2\delta_1}
           \left[
           \frac{(\frac{E_U}{\sqrt{\xi}}-\eta)e^{-\eta t}}
                {2\left( (\sqrt{\xi} + \frac{1}{\sqrt{\xi}})E_U - \eta \right)} +
           \frac{(\sqrt{\xi} + \frac{1}{\sqrt{\xi}}) \sqrt{\xi} E^2_U
                  e^{-(\sqrt{\xi} + \frac{1}{\sqrt{\xi}})E_U t} }
                {(\sqrt{\xi} + \frac{1}{\sqrt{\xi}})^2 E^2_U - \eta^2 }
           \right].
\end{equation}
As $\eta$ approaches zero, $f_D(t)$ becomes:

\begin{equation}
  f_D(t) = \delta_{\alpha\beta}\delta_{\gamma\delta}
            \frac{2\pi}{N_1^2\delta_1} \frac{1}{1+\xi} \left[ \frac{1}{2} + \xi e^{-(\sqrt{\xi} +
               \frac{1}{\sqrt{\xi}}) E_U t} \right].
\end{equation}

The full Fourier transformation includes integration over $E$ as well. In current approximation, when $E$ is close to
the center of the band, $\langle G^R_1 G^A_1 \rangle$ is independent of $E$. It will depend on $E$ if we integrate over the whole
bandwidth. The exact dependence of $\langle G^R_1 G^A_1 \rangle$ on $E$ far from the center of the band is not known. To get
correct expression we assume that integration over $E$ adds to $\langle G^R_1 G^A_1 \rangle$ multiplicative factor $N_1\delta_1$
along with normalization coefficient $A$. Also, for index pairing $\alpha = \beta$ and $\gamma =
\delta$, $G^R_{\alpha\gamma} G^A_{\delta\beta}$ becomes transition probability density $P(t)_{\alpha \rightarrow \gamma}$. 
Using equipartition
theorem, for $t \rightarrow \infty$ summation of $P(t)_{\alpha \rightarrow \gamma}$ over $\alpha$ 
one can get total probability to stay in dot 1. It is equal to $N_1/(N_1+N_2)$. That is,

\begin{equation}
  \sum_{\alpha} \int dE f_D(t) = \frac{N_1}{N_1+N_2} = \frac{1}{1+\xi}.
\end{equation}

Integration over $E$ and summation over $\alpha$ gives the factor of $A N_1^2\delta_1$. 
We identify the normalization
constant as $A = 1/\pi$. Note, that we did not use cooperon part $f_C(t)$ to determine normalization constant $A$. The reason for that is chosen index pairing. After the summation over $\alpha$ cooperon part 
contribution is of the order $1/N_1$
compared with the diffuson part. After integration over $E$ with proper normalization $f_D(t)$ 
becomes:

\begin{equation}
  f_D(t) = \delta_{\alpha\beta}\delta_{\gamma\delta}
            \frac{2}{N_1(1+\xi)} \left[ \frac{1}{2} + \xi e^{-(\sqrt{\xi} +
               \frac{1}{\sqrt{\xi}}) E_U t} \right].
\end{equation}

Then we perform the Fourier transform of the cooperon part:

\begin{equation}
  f_C(\omega) = \frac{2\pi}{N_1^2\delta_1} \frac{\delta_{\alpha\delta} \delta_{\gamma\beta}}{-i\omega}
                \frac{1 - \frac{E_{X_2}}{i\omega} - \frac{1}{\sqrt{\xi}}\frac{E_U}{i\omega} }
                     {1 - \frac{E_{X_1}+E_{X_2}}{i\omega} +
                          \frac{E_{X_1} E_{X_2}}{(i\omega)^2} +
                          \frac{E_{X_1} E_U}{\sqrt{\xi} (i\omega)^2} +
                          \frac{\sqrt{\xi} E_{X_2} E_U}{(i\omega)^2} +
                          \left( \sqrt{\xi} + \frac{1}{\sqrt{\xi}} \right)
                          \frac{E_U}{i\omega}
                          \left( \frac{E_{\Gamma}}{i\omega} - 1 \right)}.
\label{eq:a3}
\end{equation}

  The $f_C(\omega)$ is a regular function when $\omega$ approaches limiting values,
  provided at least one of the crossover energy scales $E_{X_1}$, $E_{X_2}$, or $E_{\Gamma}$ differs from zero.

To make $f_C(\omega)$ more suitable for the Fourier transform we manipulate Eq. \eqref{eq:a3} into:
\begin{equation}
  \begin{split}
  f_C(\omega) &= -\delta_{\alpha\delta} \delta_{\gamma\beta}
                 \frac{2\pi}{N_1^2\delta_1}
                 \left[ i\omega - E_{X_2} - \frac{E_U}{\sqrt{\xi}} \right] \\
          &\times \biggl[ (i\omega)^2 - \left( (E_{X_1} + \sqrt{\xi} E_U) +
                           (E_{X_2} + \frac{E_U}{\sqrt{\xi}})  \right) (i\omega) \\
                     &\qquad\qquad+\left( E_{X_1} E_{X_2} + \frac{E_{X_1} E_U}{\sqrt{\xi}} +
                        E_{X_2}E_U\sqrt{\xi} + (\sqrt{\xi}+\frac{1}{\sqrt{\xi}})E_U E_{\Gamma} 
\right)\biggr]^{-1}.
  \end{split}
\end{equation}
and observe that the poles of $f_C(\omega)$ are given by

\begin{equation}
  i\omega_{\pm} =
        \frac{(E_{X_1} + \sqrt{\xi} E_U) + (E_{X_2} + \frac{E_U}{\sqrt{\xi}}) \pm \sqrt{\mathfrak{D}}}{2}
\end{equation}
  with $\mathfrak{D} = ( (E_{X_1} + \sqrt{\xi} E_U) - (E_{X_2} + E_U/\sqrt{\xi}) )^2 +
      4 E_U^2 ( 1-4\Gamma^2 )$.
The parameter $\mathfrak{D}$ is always positive and $\omega_{\pm}$ are imaginary complex numbers.

It can be proved that
$(E_{X_1} + \sqrt{\xi} E_U) + (E_{X_2} + E_U/\sqrt{\xi}) > \sqrt{\mathfrak{D}}$ for all values of parameters, which
means that the poles are pure imaginary numbers in lower half complex plane:

\begin{equation}
  \omega_{\pm} = -i\frac{(E_{X_1} + \sqrt{\xi} E_U) + (E_{X_2} + \frac{E_U}{\sqrt{\xi}}) \pm \sqrt{\mathfrak{D}}}{2}
               = -i a_{\pm}, \hspace{0.7cm} a_+ > a_- > 0.
\label{eq:10}
\end{equation}

The function $f_C(\omega)$ now reads:

\begin{equation}
  f_C(\omega) = -\delta_{\alpha\delta} \delta_{\gamma\beta}
                 \frac{2\pi}{N_1^2\delta_1}
                 \frac{i\omega - E_{X_2} - \frac{E_U}{\sqrt{\xi}} }
                       { (i\omega - a_-)(i\omega - a_+) }.
\end{equation}

We perform the Fourier transform and use the normalization factor to obtain:

\begin{equation}
  f_C(t) = \frac{1}{2\pi} \int_{-\infty}^{+\infty} \exp(-i\omega t) f_C(\omega) d\omega dE =
     \delta_{\alpha\delta} \delta_{\gamma\beta}
     \frac{2}{N_1}
       \left[
           1 + \frac{ E_{X_2} + \frac{E_U}{\sqrt{\xi}} }{a_+ - a_-}
           \left( e^{-t a_{-}} - e^{-t a_{+}} \right)
       \right].
\end{equation}

Hence, the full expression for the Fourier transform for the two-particle Green's function in the dot are given by:

\begin{multline}
  \langle G^R_{\alpha\gamma}(t) G^A_{\delta\beta}(t)\rangle_{11} =
        \delta_{\alpha\beta} \delta_{\delta\gamma}   \frac{2}{N_1(1+\xi)}
    \left[ \frac{1}{2} + \xi e^{-(\sqrt{\xi} +
           \frac{1}{\sqrt{\xi}}) E_U t} \right] \\
  + \delta_{\alpha\delta} \delta_{\gamma\beta} \frac{2}{N_1}
       \left[
           1 + \frac{ E_{X_2} + \frac{E_U}{\sqrt{\xi}} }{a_+ - a_-}
           \left( e^{-t a_{-}} - e^{-t a_{+}} \right)
       \right]
\end{multline}
\begin{multline}
  \langle G^R_{ik}(t) G^A_{lj}(t)\rangle_{22} =
           \delta_{ij} \delta_{kl} \frac{2\xi}{N_2(1+\xi)}
  \left[
     \frac{1}{2} + \frac{1}{\xi} e^{-(\sqrt{\xi} +
               \frac{1}{\sqrt{\xi}}) E_U t} \right] \\
    + \delta_{il} \delta_{kj}\frac{2}{N_2}
       \left[
           1 + \frac{ E_{X_1} + \sqrt{\xi}E_U }{a_+ - a_-}
           \left( e^{-t a_{-}} - e^{-t a_{+}} \right)
       \right],
\end{multline}
   where $a_{\pm}$ is defined through Eq. \eqref{eq:10}.
%
%

\section{Correlation of four wave functions}\label{apnx:D}

  In this appendix we obtain correlation of four wave functions
$\langle \psi_n(\alpha) \psi^{\ast}_n(\gamma) \psi_m(\delta)
\psi^{\ast}_m(\beta) \rangle$ for the system of two coupled dots. This
has been obtained in a single dot for the pure ensembles by
supersymmetry methods by Mirlin\cite{mirlin_00}, and for the
GOE$\to$GUE crossover by Adam {\it et al}\cite{adam:article}.
  We consider ensemble average of the following product:

\begin{multline}
  \langle \left[ G^R_{\alpha\gamma}(E+\omega) - G^A_{\alpha\gamma}(E+\omega) \right]  
          \left[ G^R_{\delta\beta}(E) - G^A_{\delta\beta}(E) \right]
  \rangle 
  \approx \\
          - \langle G^R_{\alpha\gamma}(E+\omega) G^A_{\delta\beta}(E)  
          - \langle G^A_{\alpha\gamma}(E+\omega) G^R_{\delta\beta}(E)
  \rangle
  = -2(\delta_{\alpha\beta} \delta_{\gamma\delta} Re[D_1] + 
       \delta_{\alpha\delta} \delta_{\gamma\beta} Re[C_1] ),
\label{eq:20}
\end{multline}
  where $D_1$ and $C_1$ are the diffuson and cooperon expressions from Eq. \eqref{eq:18}. Here we used 
the fact that ensemble average of $G^RG^R$ and $G^AG^A$ are smaller than $G^RG^A$ and $G^AG^R$.

On the other hand, we have:

\begin{equation}
  G^R_{\alpha\gamma}(E) - G^A_{\alpha\gamma}(E) =
                         - 2\pi i\sum_n \psi_n(\alpha) \psi^{\ast}_n(\gamma) \delta(E-E_n),
\end{equation}
  and

\begin{multline}
  \langle \left[ G^R_{\alpha\gamma}(E+\omega) - G^A_{\alpha\gamma}(E+\omega) \right]
          \left[ G^R_{\delta\beta}(E) - G^A_{\delta\beta}(E) \right]
  \rangle
  \approx \\
    -4\pi^2 \langle \sum_{n,m} \psi_n(\alpha) \psi^{\ast}_n(\gamma) \psi_m(\delta) \psi^{\ast}_m(\beta)
       \delta(E + \omega - E_n) \delta(E - E_m) \rangle.
\label{eq:22}
\end{multline}

  We know that in the crossover components of eigenvalues and eigenvectors are correlated with
each other. This correlation is small already on the distances of a few $\delta$ and can be 
neglected in the limit $\omega \gg \delta$, so Eq. \eqref{eq:22} can be approximated by:

\begin{equation*}
     -4\pi^2 \langle \psi_{\bar{n}}(\alpha) \psi^{\ast}_{\bar{n}}(\gamma) 
                     \psi_{\bar{m}}(\delta) \psi^{\ast}_{\bar{n}}(\beta) \rangle
       \langle \sum_n \delta(E + \omega - E_n) \rangle 
       \langle \sum_m \delta(E - E_m) \rangle.
\end{equation*}
  where $\bar{n}$ and $\bar{m}$ mark energy levels close to $E+\omega$ and $E$ respectively.

The average of the sum is a density of states 
$\rho(E) = \langle \sum_n \delta(E-E_n) \rangle = 1/\delta$. Then, we get

\begin{multline}
    \langle \left[ G^R_{\alpha\gamma}(E+\omega) - G^A_{\alpha\gamma}(E+\omega) \right]
          \left[ G^R_{\delta\beta}(E) - G^A_{\delta\beta}(E) \right]
  \rangle
  \approx \\
  - \frac{4\pi^2}{\delta^2} 
      \langle \psi_n(\alpha) \psi^{\ast}_n(\gamma) \psi_m(\delta) \psi^{\ast}_m(\beta) \rangle.
\label{eq:21}
\end{multline}

   For the two coupled dots we have:

\begin{equation}
  Re[D_1] = \frac{2\pi}{N_1\delta_1} 
            \frac{\sqrt{\xi}E_U}{\omega^2 + (\sqrt{\xi} + \frac{1}{\sqrt{\xi}})^2 E^2_U }
\end{equation}

   In order to calculate $Re[C_1]$ from Eq. \eqref{eq:18} we are going to assume that magnetic field is zero in 
the first dot and in the hopping region ($E_{X_1} = E_{\Gamma} = 0$), and the second dot is in GOE to GUE
crossover ($E_{X_2} \sim \omega$). Then,

\begin{equation}
  Re[C_1] = \frac{2\pi}{N_1^2\delta_1}
            \frac{\sqrt{\xi}E_U \omega^2 + (E_U + \sqrt{\xi}E_{X_2})E_UE_{X_2} }
                 {(\omega^2 - \sqrt{\xi}E_UE_{X_2})^2 + (E_{X_2} + (\sqrt{\xi} 
                    + \frac{1}{\sqrt{\xi}})E_U)^2 \omega^2 }
\end{equation}

  The relation between the mean level spacing $\delta$ for the system of coupled dots and 
the mean level spacing
in the first uncoupled dot $\delta_1$ is as follows. The averaged density of states in coupled system
is going to be the sum of densities of each dot: $\langle \rho \rangle = \langle \rho_1 \rangle + 
\langle \rho_2 \rangle$, or $\delta^{-1} = \delta_1^{-1} + \delta_2^{-1}$. 
Thus, we conclude that
$\delta = \delta_1/(1+\xi)$.

   Finally, we set Eq. \eqref{eq:20} and Eq. \eqref{eq:21} equal and obtain 
correlation of for the wave functions:

\begin{multline}
  \langle \psi_n(\alpha) \psi^{\ast}_n(\gamma) \psi_m(\delta) \psi^{\ast}_m(\beta) \rangle = 
  \delta_{\alpha\beta}\delta_{\gamma\delta}\, \frac{\delta_1}{\pi (1+\xi)^2 N_1^2}\,
    \frac{\sqrt{\xi} E_U}{\omega^2 + (\sqrt{\xi} + \frac{1}{\sqrt{\xi}})^2 E_U^2 } \\
  + \delta_{\alpha\delta}\delta_{\gamma\beta}\, \frac{\delta_1 E_U}{\pi (1+\xi)^2 N_1^2}\,
            \frac{\sqrt{\xi} \omega^2 + (\sqrt{\xi}E^2_{X_2} + E_U E_{X_2}) }
                 {(\omega^2 - \sqrt{\xi}E_UE_{X_2})^2 + (E_{X_2} + (\sqrt{\xi}
                    + \frac{1}{\sqrt{\xi}})E_U)^2 \omega^2 }.
\end{multline}


\section{Sum rule for double dot system}\label{apnx:E}

   To verify the expressions we have obtained for the averaged
   Green's functions we use a sum rule.

The pair annihilation (creation) operator $T (T^{\dagger})$ in the basis of two uncoupled 
dots is a sum of two terms belonging to each dot:

\begin{equation}
  \begin{split}
  T &= \sum_{\alpha_0} c_{\alpha_0,\downarrow} c_{\alpha_0,\uparrow}
    + \sum_{i_0} c_{i_0,\downarrow} c_{i_0,\uparrow}, \\
  T^{\dagger} &= \sum_{\alpha_0} c^{\dagger}_{\alpha_0,\uparrow} c^{\dagger}_{\alpha_0,\downarrow}
    + \sum_{i_0} c^{\dagger}_{i_0,\uparrow} c^{\dagger}_{i_0,\downarrow}.
  \end{split}
\end{equation}

  Greek indices go over the states in the first dot, and Latin indices go over the states in 
the second dot. The subindex $0$ denotes the basis of two uncoupled dots.

   Our first goal is to calculate the commutator $[T^{\dagger},T]$. As operators from different dots anticommute,
one gets:
\begin{equation}
  [T^{\dagger},T] = \sum_{\alpha_0,\beta_0} 
                    [c^{\dagger}_{\alpha_0,\uparrow} c^{\dagger}_{\alpha_0,\downarrow},
                     c_{\beta_0,\downarrow} c_{\beta_0,\uparrow}]
                  +
                    \sum_{i_0,j_0}
                    [c^{\dagger}_{i_0,\uparrow} c^{\dagger}_{i_0,\downarrow},
                     c_{j_0,\downarrow} c_{j_0,\uparrow}] 
                  = 
                    \hat{N}_{1e} + \hat{N}_{2e} - N_1 - N_2,
\end{equation}
  where $\hat{N}_{1e}, \hat{N}_{2e}$ are the operators of total number of electrons
in dot 1 and dot 2, and $N_1, N_2$ are the total number of levels in dot 1 and dot 2.

    The expectation value of $[T^{\dagger},T]$ in ground state at zero temperature is:

\begin{equation}
  \overline{[T^{\dagger},T]} = \langle\Omega\vert [T^{\dagger},T] \vert\Omega\rangle = N_e - N.
\end{equation}
   $N_e$ and $N$ are the total number of electrons and levels in both dots. 
This number is conserved when going to another basis.

   Now we choose the basis of the system of coupled dots. In this basis
$c_{\alpha_0,s} = \sum_m \psi_m (\alpha_0) c_{m,s}$, and 
$c_{i_0,s} = \sum_m \psi_m (i_0) c_{m,s}$, where $c_{m,s}$ is annihilation operator in new 
basis. Using this transformation, we rewrite pair destruction operator as follows:

\begin{equation}
  T = \sum_{\alpha_0} c_{\alpha_0,\downarrow} c_{\alpha_0,\uparrow}
    + \sum_{i_0} c_{i_0,\downarrow} c_{i_0,\uparrow}
    =
      \sum_{m_1,m_2} D_{m_1 m_2} c_{m_1,\downarrow} c_{m_2,\uparrow},
\end{equation}
  where $D_{m_1 m_2}$ is defined by the following expression:

\begin{multline}
  D_{m_1 m_2} = \sum_{m_1,m_2} \left( 
                  \sum_{\alpha_0} \psi_{m_1}(\alpha_0) \psi_{m_2}(\alpha_0)   
              +
                  \sum_{i_0} \psi_{m_1}(i_0) \psi_{m_2}(i_0)
                               \right) c_{m_1,\downarrow} c_{m_2,\uparrow} \\
              =
                \sum_{p_0} \psi_{m_1}(p_0) \psi_{m_2}(p_0).
\end{multline}
   The index $p_0$ runs over all states in the first and second dots for the basis of
uncoupled dots.

In the new basis the $T,T^{\dagger}$ operators look like this:

\begin{equation}
  \begin{split}
  T &= \sum_{m_1,m_2} D_{m_1 m_2} c_{m_1,\downarrow} c_{m_2,\uparrow}, \\
  T^{\dagger} &= \sum_{m_1,m_2} D^{\ast}_{m_1 m_2} 
                    c^{\dagger}_{m_2,\uparrow} c^{\dagger}_{m_1,\downarrow}.
  \end{split}
\end{equation}

   Consequently, in the new basis,

\begin{multline}
  [T^{\dagger},T] = \sum_{m_1,m_2} \sum_{m_3,m_4} D^{\ast}_{m_1 m_2} D_{m_3 m_4}
                      [c^{\dagger}_{m_2,\uparrow} c^{\dagger}_{m_1,\downarrow},
                       c_{m_3,\downarrow} c_{m_4,\uparrow}] \\
                  =
                    \sum_{m_2,m_4} \left( \sum_{m_1} D^{\ast}_{m_1 m_2} D_{m_1 m_4}
                                   \right) c^{\dagger}_{m_2,\uparrow} c_{m_4,\uparrow}
                  -
                    \sum_{m_1,m_3} \left( \sum_{m_2} D^{\ast}_{m_1 m_2} D_{m_3 m_2}
                                   \right) c_{m_3,\downarrow} c^{\dagger}_{m_1,\downarrow}.
\end{multline}

\vspace{0.5cm}

   One can go further and use completeness condition $\sum_m \psi^{\ast}_m (p_0) \psi_m (n_0) = 
 \delta_{p_0 n_0}$ to show that in the new basis the value of commutator is
 $\hat{N}_e - N$. Our next goal, however, is to take the disorder average of the vacuum 
expectation value and to prove the invariance of $[T^{\dagger},T]$. 

   Taking into account that
$\langle\Omega\vert c^{\dagger}_{m_1,\uparrow} c_{m_2,\uparrow} \vert\Omega\rangle = 
 \delta_{m_1 m_2} \Theta(\mu - E_{m_1})$ 
and
$\langle\Omega\vert c_{m_2,\downarrow} c^{\dagger}_{m_1,\downarrow} \vert\Omega\rangle =
 \delta_{m_1 m_2} (1 - \Theta(\mu - E_{m_1}) )$, the ground state expectation value for the 
commutator is:

\begin{equation}
  \overline{[T^{\dagger},T]} = \langle\Omega\vert [T^{\dagger},T] \vert\Omega\rangle = 
               \sum_{m_1,m_2} \vert D_{m_1 m_2}\vert ^2 [2\Theta(\mu - E_{m_1}) - 1],
\end{equation}
  where $\Theta(x)$ is a step function.

   Averaging over disorder gives:

\begin{equation}
  \langle  \overline{[T^{\dagger},T]} \rangle 
         = 
           2\sum_{m_1,m_2} \Theta(\mu - E_{m_1})\langle \vert D_{m_1 m_2}\vert ^2 \rangle 
         - 
           \sum_{m_1,m_2} \langle \vert D_{m_1 m_2}\vert ^2 \rangle
\end{equation}
  Converting this into integral, we get:
\begin{multline}
  \langle  \overline{[T^{\dagger},T]} \rangle
         =
           2\int^{\mu}_{-W} \int^{W}_{-W} dE_1 dE_2 \rho(E_1)\rho(E_2) 
             \langle \vert D(E_1,E_2)\vert ^2 \rangle \\
         -
            \int^{W}_{-W} \int^{W}_{-W} dE_1 dE_2 \rho(E_1)\rho(E_2)
             \langle \vert D(E_1,E_2)\vert ^2 \rangle
  \label{eq:26}
\end{multline}
  The density of states $\rho(E)$ is the Winger's semicircle law:

\begin{equation*}
  \rho(E) = \frac{2N}{\pi W^2} \sqrt{W^2 - E^2},
\end{equation*}
  where $2W$ is the bandwidth and $N$ is the number of states in the system.

To proceed we need to find the ensemble average of the following object:

\begin{equation}
  \langle \vert D_{m_1 m_2}\vert ^2 \rangle = \sum_{p_0,n_0}
     \langle
       \psi^{\ast}_{m_1}(p_0) \psi^{\ast}_{m_2}(p_0) \psi_{m_1}(n_0) \psi_{m_2}(n_0)
     \rangle.
\end{equation}

  Using results of appendix \ref{apnx:D} one can obtain expression for the correlation
of four wave functions in the form:

\begin{multline}
  \langle
    \psi^{\ast}_{m_1}(p_0) \psi^{\ast}_{m_2}(p_0) \psi_{m_1}(n_0) \psi_{m_2}(n_0)
  \rangle
     =
       \frac{1}{2\pi^2 \rho(E_1) \rho(E_2)} 
         Re \Big[ 
              \sum_{p_0 n_0} \langle G^R_{n_0 p_0}(E_2) G^A_{n_0 p_0}(E_1) \rangle \\
          -
              \sum_{p_0 n_0} \langle G^R_{n_0 p_0}(E_2) G^R_{n_0 p_0}(E_1) \rangle
            \Big].
  \label{eq:23}
\end{multline}

   Note, that to get the correct answer for the sum rule one should keep 
$\langle G^R G^R \rangle$ term as well. Summation in Eq. \eqref{eq:23} is performed over the states 
in both dots.

   When the dots have equal mean level spacing $\delta_1 = \delta_2 = \delta_0$, one 
particle Green's function can be found exactly from the system \eqref{eq:5} without 
approximation in $U$:

\begin{equation}
  \begin{split}
  \langle G^R_{p_0 p^{'}_0}(E)\rangle = \frac{\delta_{p_0 p^{'}_0}}
                  {\frac{E}{2} + \frac{i}{2} \sqrt{W^2-E^2}}
              =
                -\frac{2i}{W} e^{i\phi} \\
  \langle G^A_{p_0 p^{'}_0}(E)\rangle = \frac{\delta_{p_0 p^{'}_0}}
                  {\frac{E}{2} - \frac{i}{2} \sqrt{W^2-E^2}}
              =
                 \frac{2i}{W} e^{-i\phi}, 
  \end{split}
\end{equation}

  where $W = 2N_0\delta_0 \sqrt{1+U}/\pi$ is the half bandwidth and $\sin\phi = E/W$. Here
both indices $p_0$ and $p^{'}_0$ belong either to the first or to the second dot.

   The sum in Eq. \eqref{eq:23} can be broken into four sums, when the indices $p_0, n_0$
belong either to the first dot, or to the second dot, or one of the indices go over the states
in the first dot, and the other one goes over the states in the second dot.

   For example, for $\langle G^R G^A \rangle$ part we have the following expression:

\begin{equation}
  \begin{split}
  \sum_{p_0 n_0} \langle G^R_{n_0 p_0} (E_2)& G^A_{n_0 p_0} (E_1) \rangle = \\
  &N_0 \left( \frac{2}{W} \right)^2 (1+U) \frac{(1+U)e^{-i\phi_{21}} - \zeta }
                {[(1+U)e^{-i\phi_{21}} - 1] [(1+U)e^{-i\phi_{21}} - \zeta] - U^2 } \\
  + &N_0 \left( \frac{2}{W} \right)^2 (1+U) \frac{(1+U)e^{-i\phi_{21}} - 1 }
         {[(1+U)e^{-i\phi_{21}} - 1] [(1+U)e^{-i\phi_{21}} - \zeta] - U^2 } \\
  + &2N_0 \left( \frac{2}{W} \right)^2 (1+U) \frac{U}
         {[(1+U)e^{-i\phi_{21}} - 1] [(1+U)e^{-i\phi_{21}} - \zeta] - U^2 }. 
  \end{split}
 \label{eq:24}
\end{equation}

\vspace{0.5cm}
\noindent  Here $\phi_{21} = \phi_2 - \phi_1$, and $\zeta = (1-X^2_{2})/(1+X^2_{2})$

   The first term in Eq. \eqref{eq:24} is the contribution of $\langle G^R \rangle \langle G^A \rangle$ plus
the cooperon part of two particle Green's function in the first dot. The second term describes 
contribution of free term and cooperon part in the second dot. The last term is a sum of transition parts 
from dot 1 to dot 2 and vice versa. It appears that these transition terms are equal, which explains 
coefficient $2$ in front of the last term in Eq. \eqref{eq:24}.

   Summation of the $\langle G^R G^R \rangle$ gives similar result:

\begin{equation}
  \begin{split}
  \sum_{p_0 n_0} \langle G^R_{n_0 p_0} (E_2)& G^R_{n_0 p_0} (E_1) \rangle = \\
 - &N_0 \left( \frac{2}{W} \right)^2 (1+U) \frac{(1+U)e^{-i\psi_{21}} + \zeta }
                {[(1+U)e^{-i\psi_{21}} + 1] [(1+U)e^{-i\psi_{21}} + \zeta] - U^2 } \\
  - &N_0 \left( \frac{2}{W} \right)^2 (1+U) \frac{(1+U)e^{-i\psi_{21}} + 1 }
         {[(1+U)e^{-i\psi_{21}} + 1] [(1+U)e^{-i\psi_{21}} + \zeta] - U^2 } \\
  + &2N_0 \left( \frac{2}{W} \right)^2 (1+U) \frac{U}
         {[(1+U)e^{-i\psi_{21}} + 1] [(1+U)e^{-i\psi_{21}} + \zeta] - U^2 },
  \end{split}
 \label{eq:25}
\end{equation}
  where $\psi_{21} = \phi_2 + \phi_1$.

   In principle, there should be terms corresponding to diffusons in dot 1 and dot 2. However, 
these terms after summation over $p_0, n_0$ are $1/N_0$ smaller than the others and in the large
$N_0$ limit can be neglected.

  Although one can use Eq. \eqref{eq:26} to verify the sum rule, it is more convenient to work with derivative of Eq. \eqref{eq:26} over $\mu$ at $\mu = 0$.

   It gives:

\begin{equation}
  \frac{\partial}{\partial\mu} \langle \overline{[T^{\dagger},T]} \rangle_{\mu = 0} 
          = 
            2\rho(0) \int_{-W}^{W} dE_2 \rho(E_2) \langle\vert D(E_1 = 0, E_2) \vert^2 \rangle.
  \label{eq:27}
\end{equation}

   On the other hand, this expression should be equal to:

\begin{equation}
  \frac{\partial}{\partial\mu} (N_e - N) = 
                   \frac{\partial}{\partial\mu} \big( 2\frac{N}{2} + 2\int_{0}^{\mu} \rho(E) dE - N \big) = 
                         2\rho(\mu).
  \label{eq:28}
\end{equation}

   Comparison of Eq. \eqref{eq:27} and \eqref{eq:28} at $\mu = 0$ results in the following condition for the sum rule:

\begin{equation}
  \int_{-W}^{W} dE_2 \rho(E_2) \langle\vert D(E_1 = 0, E_2) \vert^2 \rangle = 1.
  \label{eq:29}
\end{equation}

  The integral in Eq. \eqref{eq:29} was computed numerically and matched the unity with high accuracy.


\bibliographystyle{apsrev}
\bibliography{paper}

\end{document}